\documentclass[a4paper,fleqn,usenatbib]{mnras}

\usepackage[T1]{fontenc}
\usepackage{ae,aecompl}
\usepackage{graphicx}
\usepackage{amsmath}
\usepackage{amssymb}
\usepackage{subcaption}
\usepackage{bm}
\usepackage{newtxtext,newtxmath}
\usepackage{enumitem}
\usepackage{lastpage}
\usepackage{CJK}

\usepackage{xcolor}
\usepackage{soul}
\newcommand{\sout}[1]{\st{#1}}

\begin{document}
\begin{CJK}{UTF8}{gbsn}

\newcommand{\gsim}{\,\lower.7ex\hbox{$\;\stackrel{\textstyle>}{\sim}\;$}}
\newcommand{\lsim}{\,\lower.7ex\hbox{$\;\stackrel{\textstyle<}{\sim}\;$}}

\definecolor{darkgreen}{RGB}{0,100,0}
\newcommand{\add}[1]{{\color{darkgreen}{\bf#1}}}
\newcommand{\del}[1]{{\color{red}\sout{#1}}}
\newcommand{\delmath}[1]{{\color{red}{\ifmmode\text{\sout{\ensuremath{#1}}}\else\sout{#1}\fi}}}

\title[Impact of B-fields on thermal instability]{The impact of magnetic fields on thermal instability}
\author[Ji, Oh, McCourt]{Suoqing Ji (季索清),$^{1}$\thanks{Email: suoqing@physics.ucsb.edu} S. Peng Oh$^{1}$ and Michael McCourt$^{1,2}$ \\
$^{1}$ Department of Physics, University of California, Santa Barbara, CA 93106, USA.\\
$^{2}$ Hubble Fellow.}

\setlength{\topmargin}{-0.6cm}
\date{Accepted 0000. Received 0000; in original form 0000}

\pagerange{\pageref{firstpage}--\pageref{LastPage}}
\pubyear{2017}

\label{firstpage}
\maketitle

\begin{abstract}
Cold ($T\sim 10^{4} \ \mathrm{K}$) gas is very commonly found in both galactic and cluster halos. There is no clear consensus on its origin. Such gas could be uplifted from the central galaxy by galactic or AGN winds. Alternatively, it could form in situ by thermal instability. Fragmentation into a multi-phase medium has previously been shown in hydrodynamic simulations to take place once $t_\mathrm{cool}/t_\mathrm{ff}$, the ratio of the cooling time to the free-fall time, falls below a threshold value. Here, we use 3D plane-parallel MHD simulations to investigate the influence of magnetic fields. We find that because magnetic tension suppresses buoyant oscillations of condensing gas, it destabilizes all scales below $l_\mathrm{A}^\mathrm{cool} \sim v_\mathrm{A} t_\mathrm{cool}$, enhancing thermal instability. This effect is surprisingly independent of magnetic field orientation or cooling curve shape, and sets in even at very low magnetic field strengths. Magnetic fields critically modify both the amplitude and morphology of thermal instability, with $\delta \rho/\rho  \propto \beta^{-1/2}$, where $\beta$ is the ratio of thermal to magnetic pressure. In galactic halos, magnetic fields can render gas throughout the entire halo thermally unstable, and may be an attractive explanation for the ubiquity of cold gas, even in the halos of passive, quenched galaxies.
\end{abstract}


\section{Introduction}
\label{section:intro}

Modern theories of structure formation predict the halos of galaxies and clusters to be filled with hot virialized gas in approximate hydrostatic equilibrium (e.g, \citealt{birnboim03}). This is easily seen directly in X-ray emission at group and cluster scales, and at galaxy scales has been inferred from stacked SZ observations \citep{planck13, anderson15} and more indirect measures such as confinement of gas clouds and stripping of satellite galaxies \citep{maller04, fang13, stocke13}. In recent years, a much greater surprise from quasar absorption line observations of the circumgalactic medium (CGM) was the ubiquity and abundance of $T\sim 10^{4} \ \mathrm{K}$, photoionized gas throughout such halos, which is seen in more than half of sightlines within 100 kpc in projection \citep{werk13, stocke13}. Photoionization modeling suggests reservoirs of cold gas of $\sim 10^{10}-10^{11} \, \mathrm{M_{\odot}}$, which is $\sim 1-10\%$ of the halo mass \citep{stocke13, werk14, stern16}. At higher redshifts, when quasars are prevalent, such cold gas is also seen in area-filling, smoothly distributed fluorescent Ly$\alpha$ emission \citep{cantalupo14, hennawi15, cai17}. Importantly, deep Ly$\alpha$ imaging has found such widespread cold gas to be ubiquitous in all surveyed quasars, out to distances ranging from $\sim 100-300 \ \mathrm{kpc}$. 

Understanding the origin of this cold gas is an important challenge for several reasons. The CGM is an important discriminant of galaxy formation models, which have largely been tuned to match observed stellar masses and galaxy morphology. Furthermore, only $T\sim 10^{4} - 10^{5}\ \mathrm{K}$ gas will be observationally accessible in detail for the foreseeable future. Finally, cold gas both provides fuel for star formation and must be expelled in outflows, and thus is arguably the most important component to track and understand. 

The theoretical challenge is similar to that in galaxy cluster cores, where the observed cold filaments are thought to arise either from uplifted cold gas from the central cD galaxy, or are made in situ via thermal instability. In galaxies, models where cold gas is uplifted by galactic winds face problems in understanding how gas can be entrained without being shredded by hydrodynamic instabilities (e.g., \citealt{scannapieco15, zhang17}), leading to suggestions that such gas must instead instead spawned by thermal instability in the wind \citep{martin15, thompson16, scannapieco17}. It is also hard to understand the prevalence of dense cold gas in passive galaxies. These problems may potentially be alleviated in recent models which invoke small scale structure in cold gas to enable rapid entrainment and suspension of cold gas droplets \citep{mccourt18}. Nonetheless, thermal instability remains an attractive way to produce cold gas in situ and solve these problems. 

Simulations of local thermal instability in stratified systems in global hydrostatic and thermal balance showed that fragmentation into a multi-phase medium happens if the ratio of the local cooling time to the free-fall time falls below a threshold value of $t_\mathrm{cool}/t_\mathrm{ff} \sim 1-10$. Subsequent simulations by other groups have corroborated this finding, and have also extended its relevance to another domain, suggesting that black holes are fed by cold accretion, maintaining a feedback loop that pins $t_\mathrm{cool}/t_\mathrm{ff}$ at this threshold value \citep{gaspari12, gaspari13, li14, li15, prasad15}. Observationally, there is also support for the idea that an interplay between cooling and feedback leads cluster cores to self-regulate at a threshold value $t_\mathrm{cool}/t_\mathrm{ff} \sim 10$ \citep{sharma12, voit15}. The exact origin of this numerical value for the threshold is still debated \citep{voit17}, and the results of this paper will bear directly upon this. These ideas also lead to the notion that star formation in galaxies may be precipitation regulated \citep{voit15_galaxy}.

Observational constraints on magnetic fields in galaxy halos are relatively sparse and uncertain. In the Milky Way, best-fitting models indicate a field of $1-10 \, \mu$ G (see Table 1 and section 17.4 of \citealt{haverkorn15}). Assuming equipartition between cosmic-rays and magnetic fields, the scale height from synchrotron emission is $5-6 \ \mathrm{kpc}$ \citep{cox05}. Similarly large scale heights of $\ge 7$ kpc are deduced\footnote{Assuming equipartition, the total (ordered and irregular) B-field scale height is at least $3-\alpha$ times larger than the $\sim 2$kpc synchrotron scale height, where $\alpha \approx -0.8$ is the synchrotron spectral index \citep{beck13}.} from the radio halos of edge-on galaxies \citep{dumke98,heesen09}. Faraday rotation along radio quasar sightlines yield magnetic fields as large as tens of $\mu$G out to $\sim 50$ kpc distances from galaxies, out to $z \sim 1$ \citep{bernet13}, likely due to outflows. Faraday rotation also yields $\sim 10 \, \mu$G fields in the halo of the edge-on spiral NGC 4631 \citep{mora13}.\footnote{For more details on magnetic field observations in galaxy halos, see a recent review paper \citet{han2017observing}.}

At face value, these field strengths are far greater than the pressure of gas in the cold phase \citep{werk14}, and comparable to or possibly higher than the pressure of the ambient hot gas\footnote{This could suggest that the cold phase is supported by non-thermal pressure such as magnetic fields, turbulence or cosmic rays (see \citealt{wiener17-cold-clouds} for further discussion of the last possibility), and further highlight the importance of considering non-thermal components of the plasma.} (for which indirect measurements and upper limits suggest $P_{\text{hot}}\lesssim10^{-13}\ \mathrm{erg/cm^3}$; \citealt{fang13,anderson15}).  This is in contrast to the previously studied situation in galaxy clusters, where the magnetic field is highly subdominant: $\beta=P_\mathrm{gas}/P_\mathrm{mag}\sim50$. Such high magnetic fields may potentially arise from galactic outflows, the $\alpha-\Omega$ dynamo, or magnetic buoyancy of strong fields produced in the galaxy. Regardless, observations suggest that the gas in the outer parts of galaxy halos is magnetically dominated, and therefore that the hydrodynamic calculations in the literature may miss essential aspects of the dynamics by neglecting magnetic fields.
    
Indeed, strong magnetic fields can fundamentally change the physics of thermal instability. Local thermal instability in a stratified medium produces overstable gravity waves, whose non-linear saturation depends on $t_\mathrm{cool}/t_\mathrm{ff}$, representing the competition between the driving processes of cooling and the damping effect of buoyant oscillations. Magnetic pressure can alter gas buoyancy, while magnetic tension can suspend overdense gas. Our goal in this paper is to carefully evaluate such effects in 3D MHD simulations. Previous work has already suggested that magnetic fields can promote thermal instability \citep{loewenstein90, balbus91}.  However, there are two important distinctions between this work and that of \citet{loewenstein90} and \citet{balbus91}. Firstly, the background state they assume is a cooling flow. In the purely hydrodynamic case, thermal instability cannot develop \citep{balbus89,li2012simulating}. By contrast, we assume a background state in hydrodynamic and thermal equilibrium, where thermal instability and a multi-phase medium \emph{can} develop, if $t_\mathrm{cool}/t_\mathrm{ff}$ is sufficiently low \citep{mccourt12,sharma12}. Secondly, those papers perform a linear stability analysis. Since the damping of thermal instability by gas motions is fundamentally a non-linear process, it is only by running MHD simulations that we can find the non-linear endstate of thermal instability.

The outline of this paper is as follows. In \S\ref{sect:methods}, we describe our methods; in \S\ref{sect:results}, we describe our numerical results; and in \S\ref{sect:interpretation} we provide a physical interpretation. Finally, in \S\ref{sect:discussion} we discuss the potential impact of some missing physics in our simulations, and summarize our conclusions.  

\section{Methods}
\label{sect:methods}

  We use FLASH v4.3 \citep{fryxell00} developed by the FLASH Center of the University of Chicago for our simulations. A directionally unsplit staggered mesh (USM) MHD solver, based on a finite-volume, high-order Godunov method combined with a constrained transport (CT) type of scheme, is adopted to solve the following fundamental governing equations of inviscid ideal magnetohydrodynamics \citep{tzeferacosetal12, lee2013solution}:
  \begin{subequations}
    \begin{gather}
      \frac{\partial\rho}{\partial t} + \bm{\nabla} \cdot (\rho \bm {v}) = 0 \\
      \frac{\partial\rho \bm{v}}{\partial t} + \bm{\nabla} \cdot \left(\rho \bm{v} \bm{v} - \frac{\bm{B} \bm{B}}{4\pi}\right) + \bm{\nabla} P_\mathrm{tot} = \rho \bm{g} \\
      \rho T \left(\frac{\partial s}{\partial t} + \bm {v} \cdot \nabla s \right) = \mathcal{H} - \mathcal{L} \\
      \frac{\partial \bm{B}}{\partial t} + \bm{\nabla} \cdot (\bm{v}\bm{B} - \bm{B} \bm{v}) = 0 \\
      \nabla \cdot \bm{B} = 0
    \end{gather}
    \label{eq:MHD}
  \end{subequations}
  where $p_\mathrm{tot} = p + B^2 / (8 \pi)$ is the total pressure including both gas pressure $p$ and magnetic pressure $B^2 / (8 \pi)$, $\bm{g}$ is gravitational acceleration, $s = c_\mathrm{p} \mathrm{ln} (P \rho^{-\gamma})$ is entropy per mass, and $\mathcal{H}$ and $\mathcal{L}$ are the source terms of heating and cooling. We adopt a plane-parallel geometry with a symmetric domain about $z = 0$ plane, where the gravitational acceleration aligns vertically along the $z$ axis:
  \begin{align}
    \bm{g} = g_0 \frac{z/a}{\sqrt{1+(z/a)^2}} \hat{z}
  \end{align}
  which describes a nearly constant gravitational acceleration for $|z| > a$ with a smooth transition to zero at $z=0$. Here the softening distance is set to $1/10$ of the disk scale height. The free-fall timescale associated with the gravitational field is:
  \begin{align}
    t_\mathrm{ff} = \sqrt{\frac{2z}{g_0}}
  \end{align}
  With this gravitational acceleration we set up the initial condition in hydrostatic equilibrium, where temperature $T_0$ is constant, the density profile $\rho$ is:
  \begin{align}
    \rho(z) = \rho_0 \mathrm{exp}\left[-\frac{a}{H}\left(\sqrt{1+\frac{z^2}{a^2}} -1\right)\right],
  \end{align}
  and the scale height $H=T_0/g_0$. For convenience, we rescale our unit system by setting $\rho_0 = T_0 = g_0 = k_\mathrm{B} = \mu m_\mathrm{p} = 1$.

  We performed all of our simulations in 3D Cartesian grids, with a resolution of $256^3$ for standard runs and higher resolutions for convergence tests. The domain of the simulation is set to be $6 \times 6 \times 6$ in the units of scale height $H$. We adopt hydrostatic boundary conditions for the boundaries along $z$ direction to maintain hydrostatic equilibrium in the initial conditions when both cooling and heating are turned off, and periodic boundary conditions for other boundaries. We initialize isobaric density perturbations with a white noise power spectrum and amplitude of $1\%$ and wavenumber ranging from $1$ to $20$ in units of $2\pi / L_\mathrm{box}$ where $L_\mathrm{box}$ is the domain size. The magnetic fields are initialized in the horizontal direction with $\beta_0$ ranging from infinity to order unity. We also performed additional simulations with initially vertical fields, which are discussed in \S\ref{sect:orientation}. These straight fields are initialized to have uniform strength throughout the box, so in the initial background state the magnetic fields do not exert any forces (and do not contribute to hydrostatic equilibrium). Note that our setup implies that plasma beta falls with height in the background state. The value we quote is $\beta$ evaluated at $z = H$.

  It is well-known that halo gas in galaxies and galaxy clusters requires some form of ``feedback heating'' to prevent a global cooling catastrophe. Precisely how feedback works is still an active area of research. Hence, we will follow \citet{mccourt12} and implement a simple model which preserves thermal energy at each radius (or $z$) in the domain: to maintain global thermal balance, in each time step we sum up total amount of cooling over the volume of a thin slab perpendicular to $z$ axis, and distribute heat uniformly over this volume. In the region of $|z| \leq a$, heating and cooling are turned off due to the uncertainty of feedback prescription at small radii. This idealization allows us to focus on the physics of thermal instability. Later works (e.g. \citealt{sharma12, gaspari13}) have shown that results are not sensitive to the specific implementation of heating.

\begin{figure}
    \begin{center}
      \includegraphics[width=0.5\textwidth]{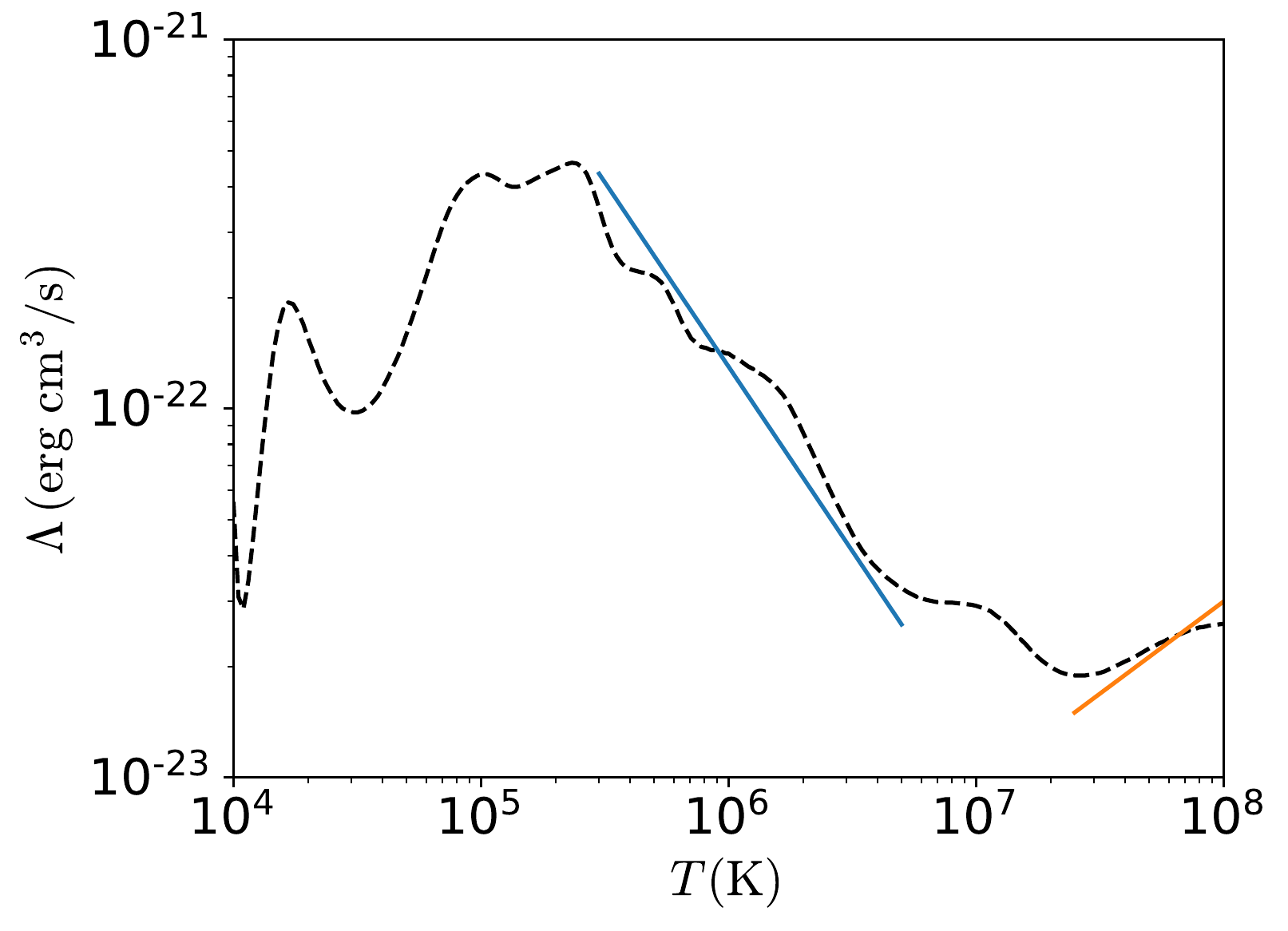}
      \caption{Approximate power-law fits for cooling curve (dashed line), with index of $-1$ for galaxy halos (blue) and $0.5$ for galaxy clusters (orange).}
      \label{fig:fit_cooling}
    \end{center}
  \end{figure}

  As an approximate fit to slopes in the regimes of heavy element excitation and bremsstrahlung respectively (Fig. \ref{fig:fit_cooling}), we implement cluster-like and galaxy-like power-law cooling functions as the following:
  \begin{align}
    \mathcal{L} = n^2 \Lambda(T) =
    \begin{cases}
      \Lambda_0 n^2 T^{1/2} \Theta (T-T_\mathrm{floor}) \qquad \text{galaxy cluster}
        \\
      \Lambda_0 n^2 T^{-1} \Theta (T-T_\mathrm{floor}) \qquad \text{galaxy halo}
    \end{cases},
    \label{eq:cool_func}
  \end{align}
  where $\Theta$ is Heaviside function. These cooling functions correspond to two cooling timescales:
  \begin{align}
    t_\mathrm{cool} =
    \begin{cases}
      \displaystyle\frac{5}{3}\frac{T^{1/2}}{n\Lambda_0} \qquad \text{galaxy cluster}
      \\
      \\
      \displaystyle\frac{5}{6}\frac{T^{2}}{n\Lambda_0} \qquad \text{galaxy halo}
    \end{cases}.
  \end{align}
The different prefactors are a result of the different cooling curve slopes (see equations 8 and 9 in \citealt{sharma12}). The $T_\mathrm{floor}$ (set to $T_0/20$) limit prevents cold gas from collapsing to very small values which we cannot resolve. Physically, our assumption is that once gas reaches this floor, its cooling time becomes so short that it will cool all the way to $T \sim 10^{4}\ K$. In addition, the temperature could reach arbitrarily large values under our simple heating prescription, which is not physical. We therefore limit the temperature to some maximum $T_\mathrm{ceiling}$ (set to $5T_0$). Physically, these limits of $T_\mathrm{cool}$ and $T_\mathrm{ceiling}$ correspond to the turning points in cooling function at which gas can reach at the two-phase semisteady state.

  The cooling time in multiphase gas can become very much shorter than the dynamical time. Computational methods which limit the time step to some fraction of a cooling time thus become prohibitively expensive. We therefore implement the ``exact'' cooling algorithm described in \citet{townsend09} into FLASH code, where we solve the operator-split energy equation
  \begin{align}
    t_\mathrm{cool} \frac{dT}{dt} = - T_\mathrm{ref} \frac{\Lambda(T)}{\Lambda(T_\mathrm{ref})}
  \end{align}
  by separating variables where $T_\mathrm{ref}$ is arbitrary reference temperature, which has an analytic solution for (piecewise) power-law cooling functions. With this implementation, we can simulate rapid cooling, even when the cooling time dips below the simulation time step.

\section{Results}
\label{sect:results}

  We perform a series of simulations with different initial values of $t_\mathrm{cool}/t_\mathrm{ff}$ and magnetic field strengths, and we plot the magnitude of density fluctuations as a function of $t_\mathrm{cool}/t_\mathrm{ff}$ in Fig. \ref{fig:rho_vs_time_cluster}. In this figure, each data point is taken from an individual simulation with the corresponding value of initial $t_\mathrm{cool}/t_\mathrm{ff}$. The magnitude of density fluctuations is computed by volumetrically averaging over the region $0.9H<|z|<1.1H$. It is defined as:
  \begin{align}
    \frac{\delta\rho}{\rho} = \frac{\rho - \langle \rho\rangle}{\langle \rho \rangle},
  \end{align}
  where the bracket represents the spatial averaging over at the same height. This quantity is computed from a snapshot of each simulation which has been evolved over $\sim 17\ t_\mathrm{cool}$, when thermal instability saturates and the computed quantity enters a stable stage. Note that in our unit system we change the cooling time by changing $\Lambda_0$; physically, however, this corresponds to changing the gas density.

\subsection{Amplitude of density fluctuations}
\label{sect:dens_fluc}

\begin{figure}
    \begin{center}
      \includegraphics[width=0.5\textwidth]{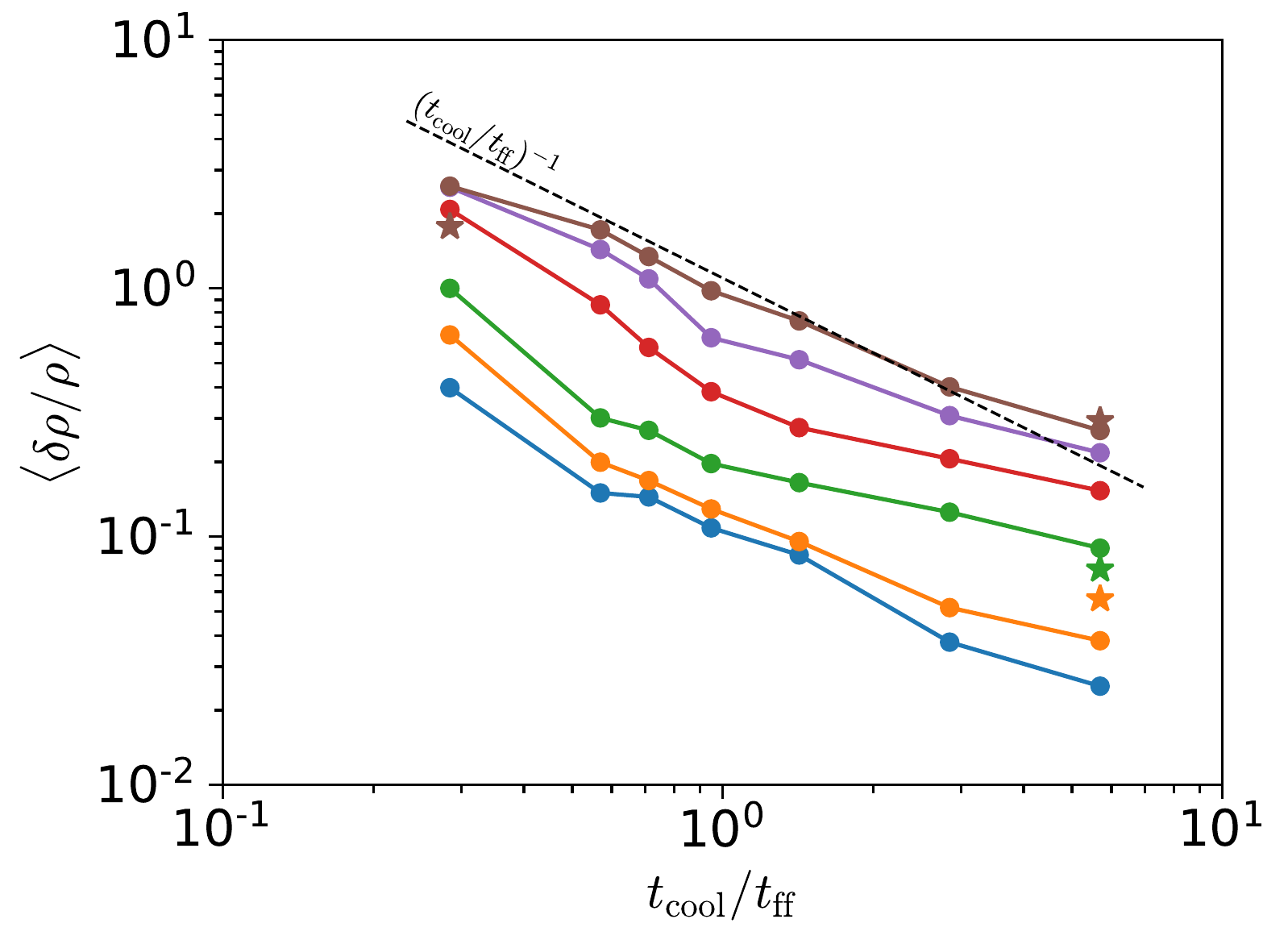}
      \caption{$\delta \rho/\rho$ as a function of $t_\mathrm{cool}/t_\mathrm{ff}$ with galaxy cluster cooling curve and different initial strengths of magnetic field characterized by the values of $\beta_0$: blue -- $\beta_0 = \infty$, orange -- $\beta_0 = 772$, green -- $\beta_0 = 278$, red -- $\beta_0=27$, purple -- $\beta_0=4$, brown -- $\beta_0=3$. The dashed lines are power-law fits to simulation data points. The dot markers are simulations with initially horizontal field, star markers are with vertical field. All the quantities are measured at $z=H$.}
      \label{fig:rho_vs_time_cluster}
    \end{center}
  \end{figure}

\begin{figure}
    \begin{center}
      \includegraphics[width=0.5\textwidth]{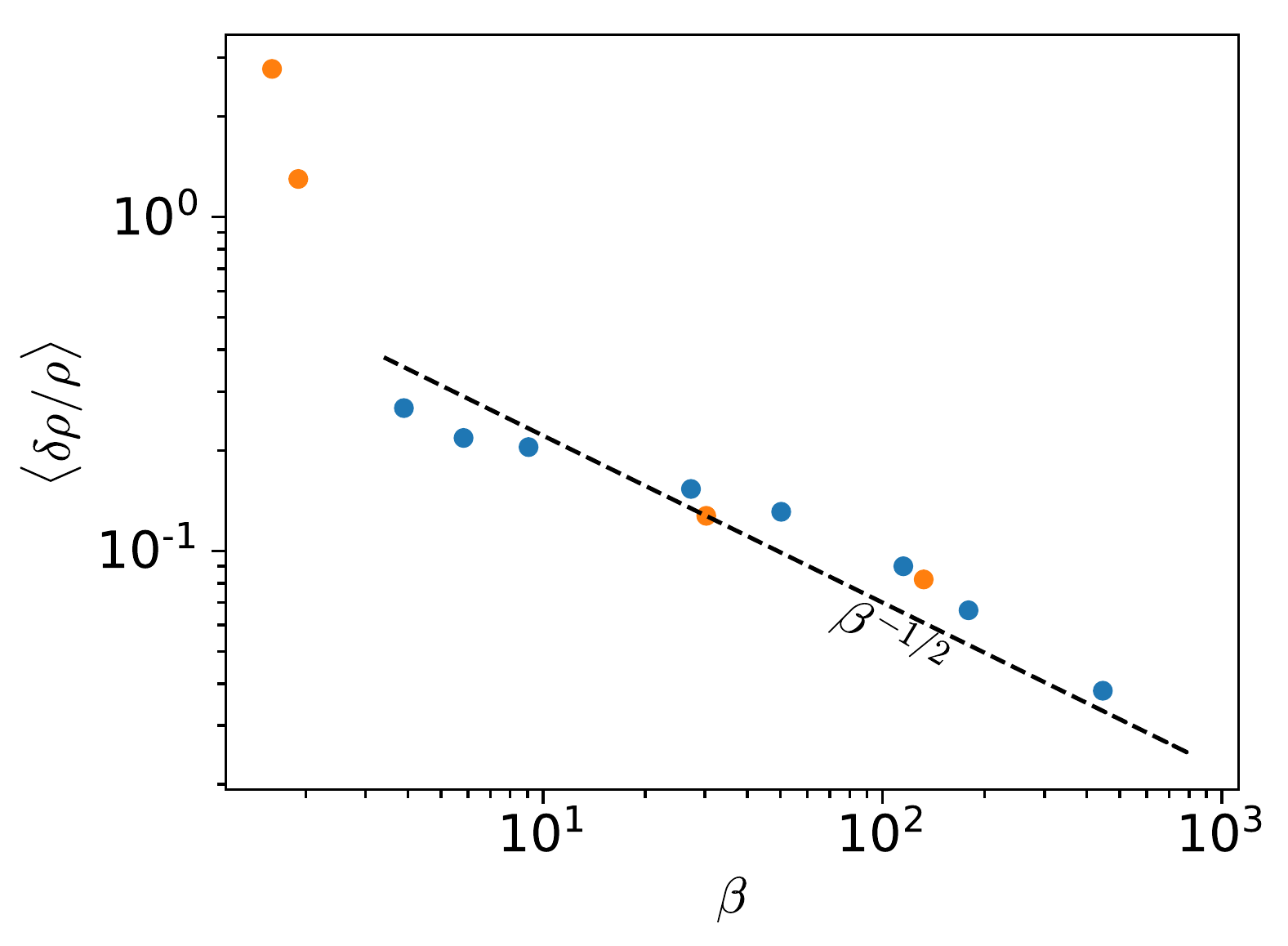}
      \caption{The $\beta$ dependence of $\delta \rho/\rho$, where both $\delta \rho/\rho$ and $\beta$ are taken from the snapshots at $t_\mathrm{cool}/t_\mathrm{ff} = 5.7$ and averaged by volume over a thin slab at $z=H$. The blue dots are with galaxy cluster cooling curve, and the orange dots are with the galaxy halo cooling curve.}
      \label{fig:rho_vs_beta}
    \end{center}
  \end{figure}

  In Fig. \ref{fig:rho_vs_time_cluster}, we show $\delta\rho/\rho$ as a function of $t_\mathrm{cool}/t_\mathrm{ff}$ for various values of magnetic field strength and a galaxy cluster-like cooling curve ($\Lambda\propto T^{-1/2}$). The hydrodynamic simulations have a $\delta\rho/\rho \propto (t_\mathrm{cool}/t_\mathrm{ff})^{-1}$ scaling, as found by \citet{mccourt12}. Two striking facts emerge from this plot. For this cooling curve, the MHD simulations retain the overall scaling $\delta\rho/\rho \propto (t_\mathrm{cool}/t_\mathrm{ff})^{-1}$, albeit with a normalization which increases as the magnetic field strength increases (i.e. plasma $\beta$ decreases). Evidently, magnetic fields promote thermal instability. In Fig. \ref{fig:rho_vs_beta}, we quantify this\footnote{We only plot this scaling for $t_\mathrm{cool}/t_\mathrm{ff} = 5.7$, but it continues to hold for other values of $t_\mathrm{cool}/t_\mathrm{ff}$.}: $\delta\rho/\rho \propto \beta^{-1/2}$ for both galaxy and cluster-like cooling curves, with a strong uptick signaling runaway cooling at low $\beta$ for galaxy-like cooling curves. Secondly, magnetic fields have an impact even when they are remarkably weak. Fig. \ref{fig:rho_vs_time_cluster} shows that the MHD case converges to the hydrodynamic case only when $\beta\sim1000$. In short, for a cluster cooling curve we find that the simulations reach a steady-state amplitude:
  \begin{align}
    \frac{\delta\rho}{\rho} \sim
    \begin{cases}
      \displaystyle 0.1 \left(\frac{t_\mathrm{cool}}{t_\mathrm{ff}}\right)^{-1} \qquad &\text{hydro}
      \\
      \\
      \displaystyle 0.3 \beta_{100}^{-1/2} \left(\frac{t_\mathrm{cool}}{t_\mathrm{ff}}\right)^{-1} \qquad &\text{MHD}
    \end{cases}.
    \label{eq:delrho_time}
  \end{align}
  While these fits are derived for horizontal fields and a cluster-like cooling curve, we shall see that (as hinted in Fig. \ref{fig:rho_vs_beta}) they are  independent of cooling curve slope and field geometry down to $\beta \gtrsim$ a few.

\begin{figure}
    \begin{center}
      \includegraphics[width=0.5\textwidth]{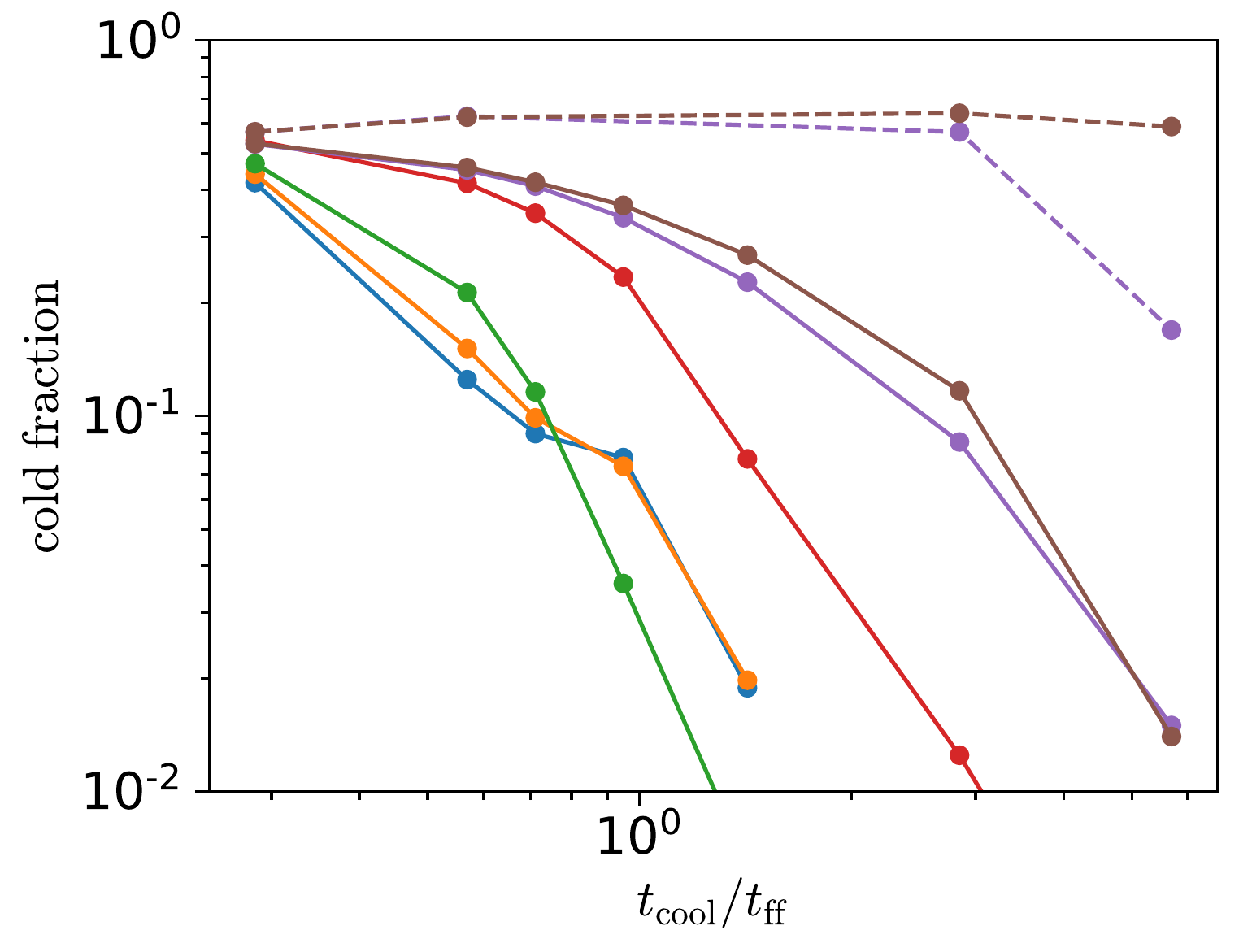}
      \caption{The mass fraction of cold gas (defined as $T < T_0/3$) over entire volume, with galaxy cluster cooling curve and different initial strengths of magnetic field characterized by the values of $\beta_0$: blue -- $\beta_0 = \infty$, orange -- $\beta_0 = 772$, green -- $\beta_0 = 278$, red -- $\beta_0=27$, purple -- $\beta_0=4$, brown -- $\beta_0=3$. The solid lines are with cluster cooling curve, and the dashed ones are with galaxy cooling curve. For $\beta_0 = \infty$ and $\beta_0 = 772$ cases, there is no detectable cold gas for simulations with $t_\mathrm{cool}/t_\mathrm{ff}\geq 2.8$.}
      \label{fig:cold_vs_t}
    \end{center}
  \end{figure}

  The amplitude of density fluctuations is important because it signals whether spatially extended multi-phase structure can develop in the saturated state of thermal instability. In Fig. \ref{fig:cold_vs_t}, we show the cold mass fraction (where $T < T_0/3$) as a function of $t_\mathrm{cool}/t_\mathrm{ff}$. For stronger magnetic fields, the threshold value of $t_\mathrm{cool}/t_\mathrm{ff}$ at which multi-phase gas can form is higher by up to an order of magnitude. Since $t_\mathrm{cool}/t_\mathrm{ff}$ varies as a function of radius, this implies that cold gas can be produced over an increasingly wider radial range for magnetized halo gas.\footnote{Strikingly, the cold fraction becomes independent of $t_\mathrm{cool}/t_\mathrm{ff}$ for low $\beta$ and a galaxy cooling curve, an effect we soon discuss.} 

\begin{figure}
    \begin{center}
      \includegraphics[width=0.5\textwidth]{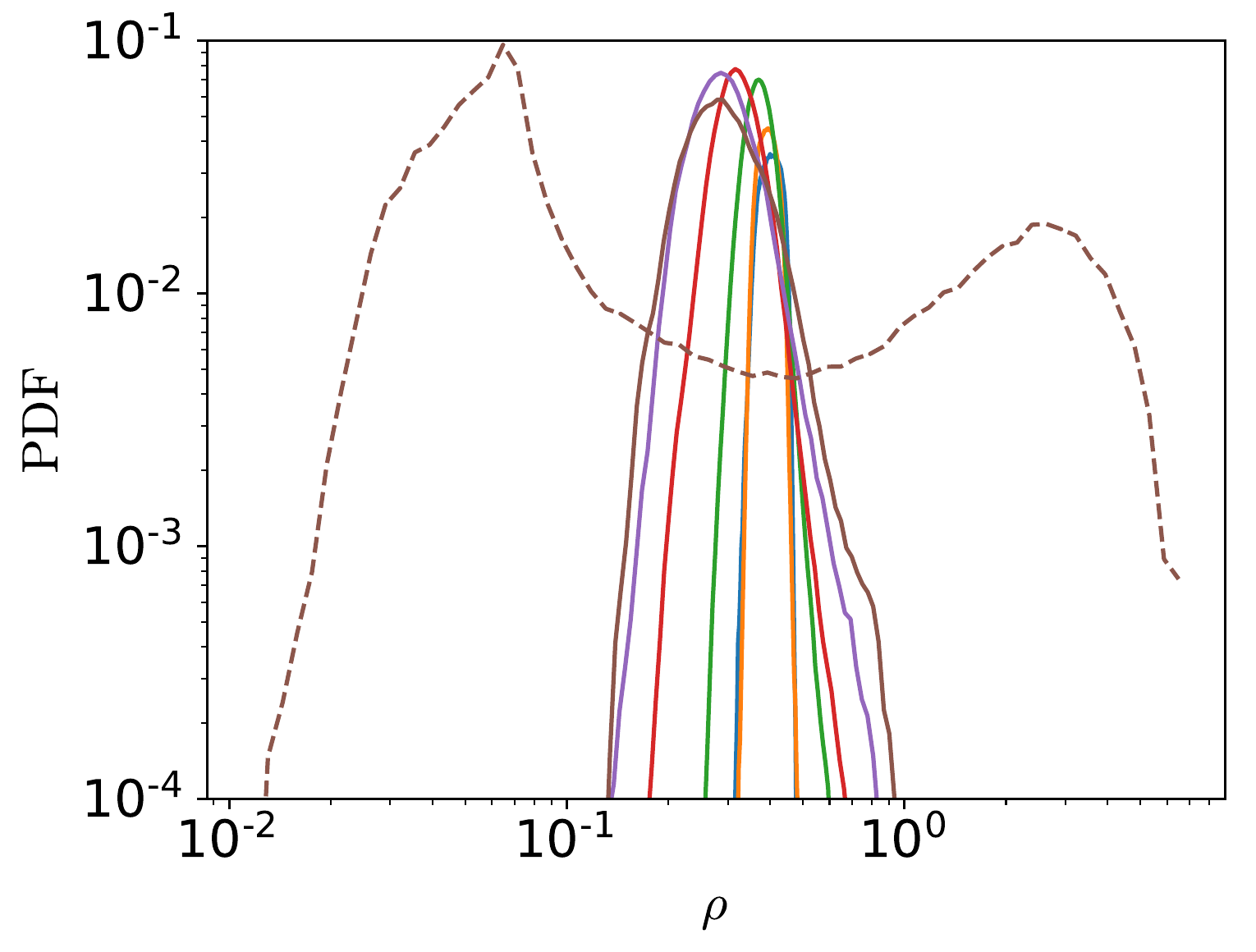}
      \caption{The probability density distribution of gas density measured at $z=H$, with galaxy cluster cooling curve and different initial strengths of magnetic field characterized by the values of $\beta_0$: blue -- $\beta_0 = \infty$, orange -- $\beta_0 = 772$, green -- $\beta_0 = 278$, red -- $\beta_0=27$, purple -- $\beta_0=4$, brown -- $\beta_0=3$. Simulations used here are with cluster cooling curve and at long cooling time ($t_\mathrm{cool}/t_\mathrm{ff} = 5.7$). Also shown with a dashed line is the $\beta_0=3$ galaxy cooling curve.}
      \label{fig:density_pdf}
    \end{center}
  \end{figure}

  In Fig \ref{fig:cold_vs_t}, the cold gas fraction apparently does not scale monotonically with $\beta$ at high $\beta$ (note the cross-over at $t_{\rm cool}/t_{\rm ff}=0.7$ between the $\beta=278$ (green), $\beta=772$, and hydro (blue) curves, in apparent contradiction with the monotonic scaling of $\delta \rho/\rho \propto \beta^{-1/2}$ seen in Fig \ref{fig:rho_vs_beta}. In fact, at higher $t_{\rm cool}/t_{\rm ff}$ (off the plot scale), there is more cold gas in the $\beta=278$ simulation than in the $\beta=772$ and hydro simulation, where cold gas is undetectable. This is a consequence of the somewhat arbitrary definition of cold gas as $T< T_{0}/3$. The gas density PDFs narrow monotonically as $\beta$ increases (Fig \ref{fig:density_pdf}), which is the more physically important measure. The cold gas mass becomes sensitive to the exact definition as the PDF narrows. Also, note that while overall trends should be robust, detailed results are sensitive to both the cooling curve shape (which we have simplified) and also our cooling curve cutoff at $T_{0}/20$. In reality, of course, the density PDF should be bimodal.

  To reinforce this point that magnetic fields can play a key role in the development of a multi-phase medium, in Fig. \ref{fig:density_pdf} we show the PDF of gas density for different values of $\beta$, for the same parameters ($z=H$, cluster cooling curve, $t_\mathrm{cool}/t_\mathrm{ff}=5.7$). Stronger magnetic fields (decreasing $\beta$) clearly broaden the gas density considerably. This is reminiscent of how decreasing values of $t_\mathrm{cool}/t_\mathrm{ff}$ broaden the gas density PDF \citep{mccourt12}, and buttresses the expectation from Fig. \ref{fig:rho_vs_time_cluster} and Eq. \eqref{eq:delrho_time} that higher values of $t_\mathrm{cool}/t_\mathrm{ff}$ can be compensated with stronger magnetic fields to produce comparable density fluctuations, even if (as we shall see) the morphology of cold gas is completely different.

\subsection{Morphology}
\label{sect:morph}

\begin{figure*}
    \begin{subfigure}[b]{0.48\textwidth}
      \centering
      \includegraphics[width=\textwidth]{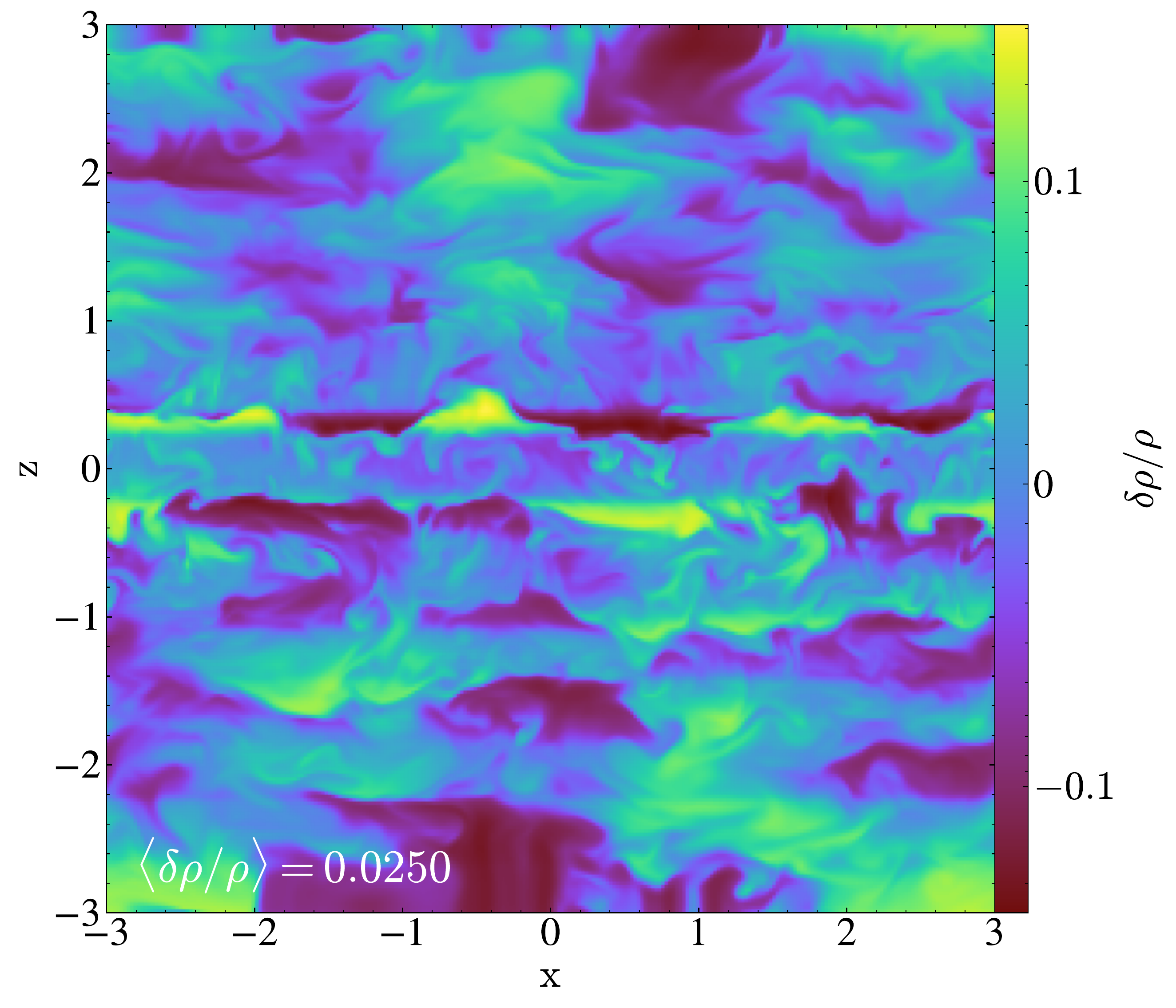}
      \caption{$\beta_0=\infty$, $t_\mathrm{cool}/t_\mathrm{ff} = 5.7$, cluster cooling curve}
      \label{fig:deltarho_betainf_cluster}
    \end{subfigure}
    \begin{subfigure}[b]{0.48\textwidth}
      \centering
      \includegraphics[width=\textwidth]{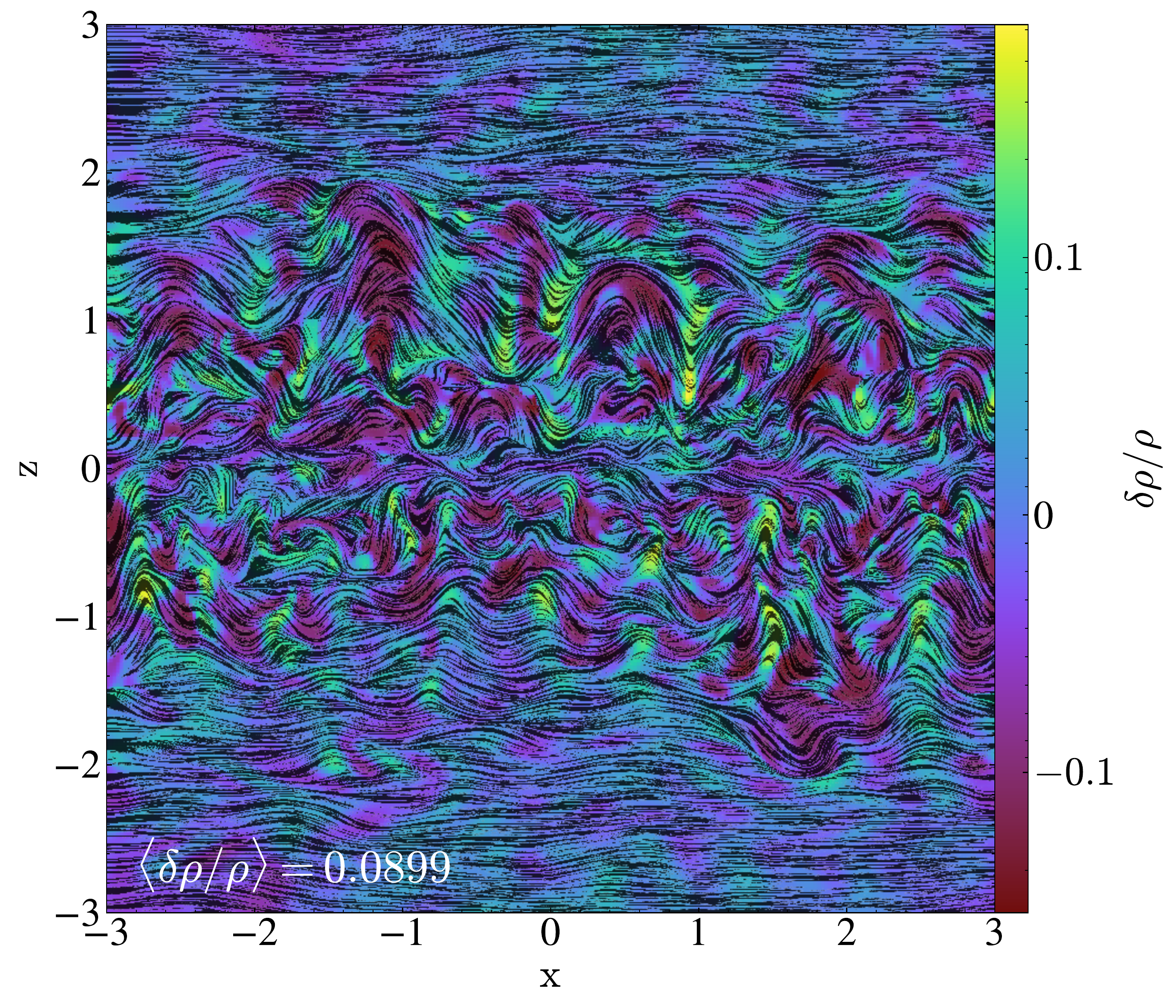}
      \caption{$\beta_0=278$, $t_\mathrm{cool}/t_\mathrm{ff} = 5.7$, cluster cooling curve}
      \label{fig:deltarho_beta278_cluster}
    \end{subfigure}
    \begin{subfigure}[b]{0.48\textwidth}
      \centering
      \includegraphics[width=\textwidth]{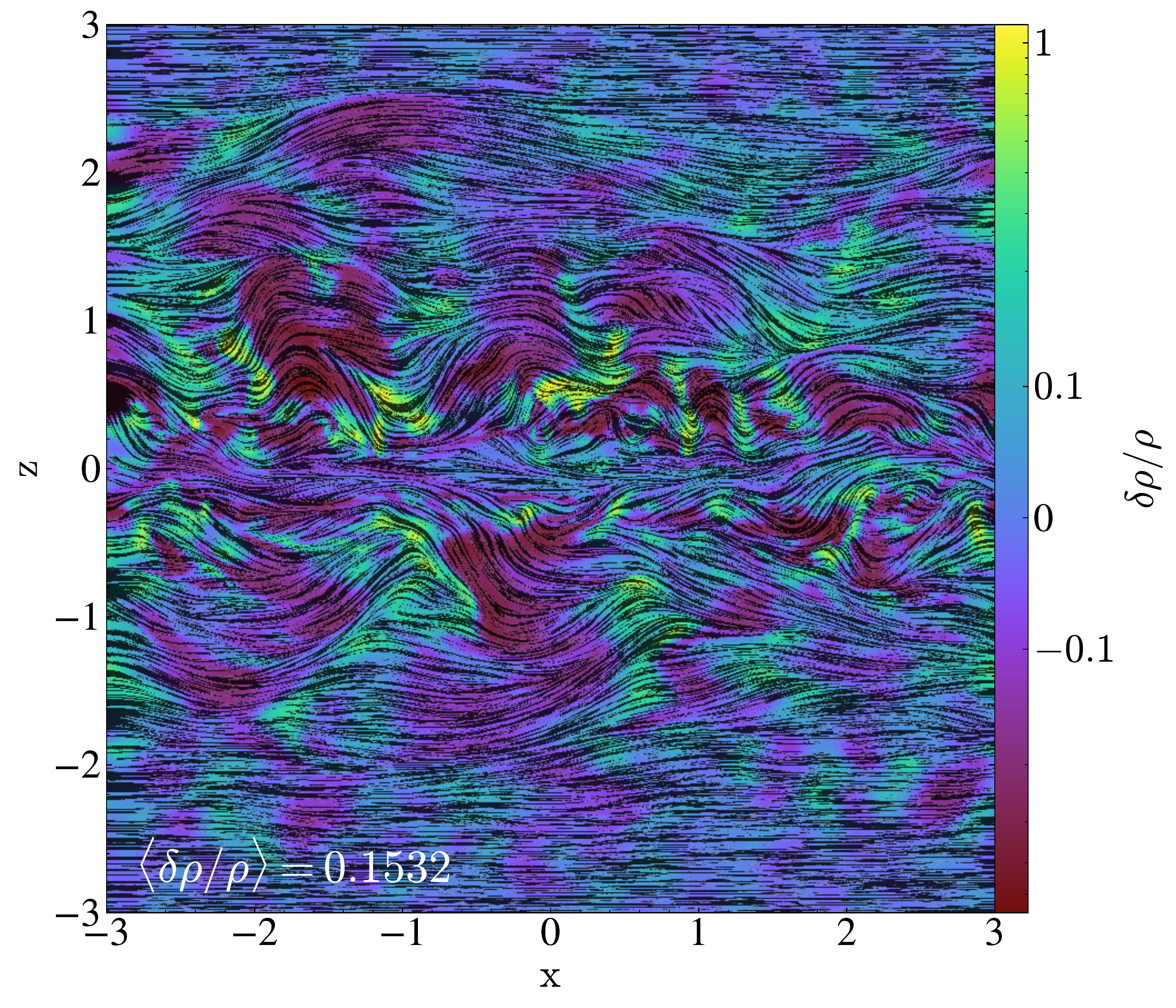}
      \caption{$\beta_0=27$, $t_\mathrm{cool}/t_\mathrm{ff} = 5.7$, cluster cooling curve}
      \label{fig:deltarho_beta27_cluster}
    \end{subfigure}
    \begin{subfigure}[b]{0.48\textwidth}
      \centering
      \includegraphics[width=\textwidth]{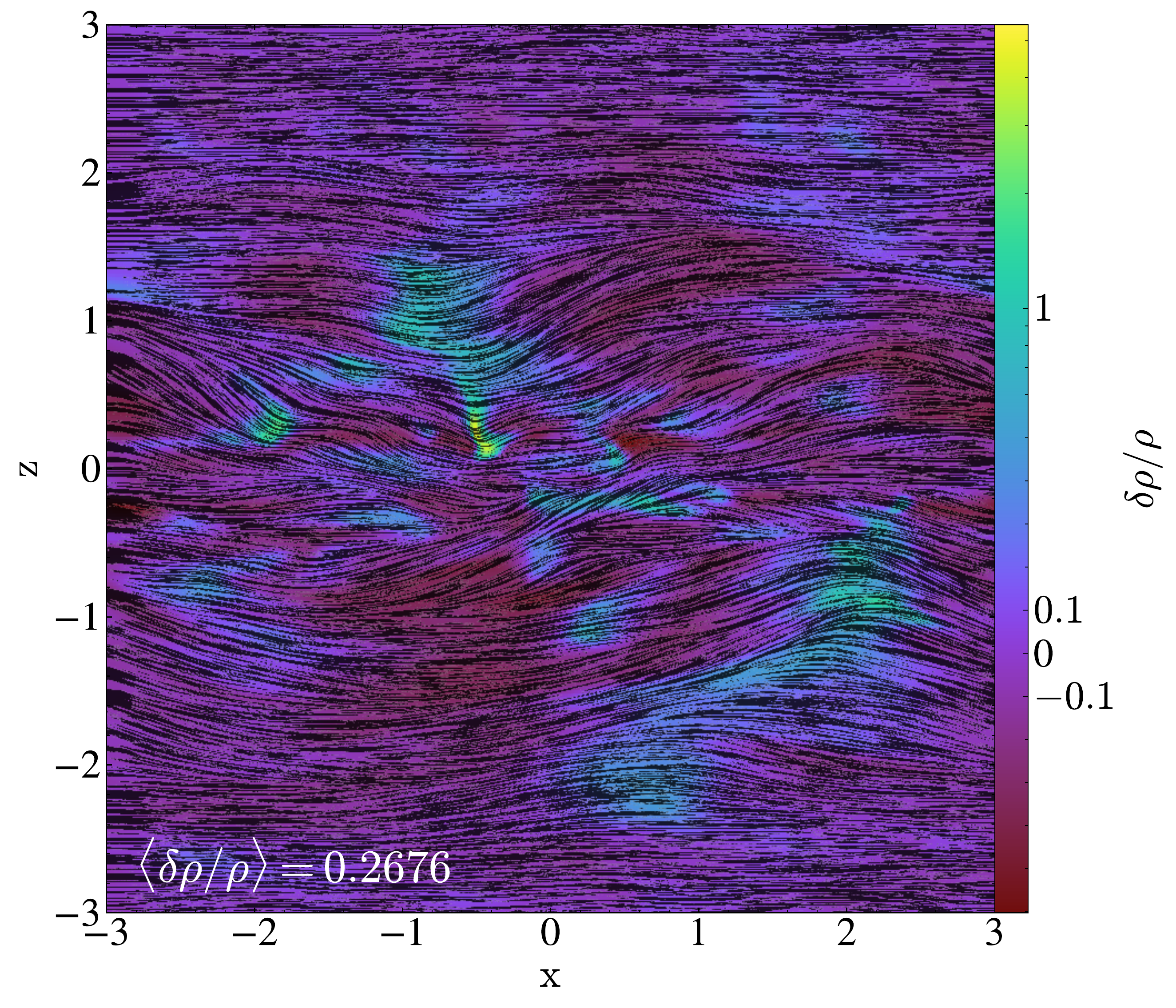}
      \caption{$\beta_0=3$, $t_\mathrm{cool}/t_\mathrm{ff} = 5.7$, cluster cooling curve}
      \label{fig:deltarho_beta3_cluster}
    \end{subfigure}
    \begin{subfigure}[b]{0.48\textwidth}
      \centering
      \includegraphics[width=\textwidth]{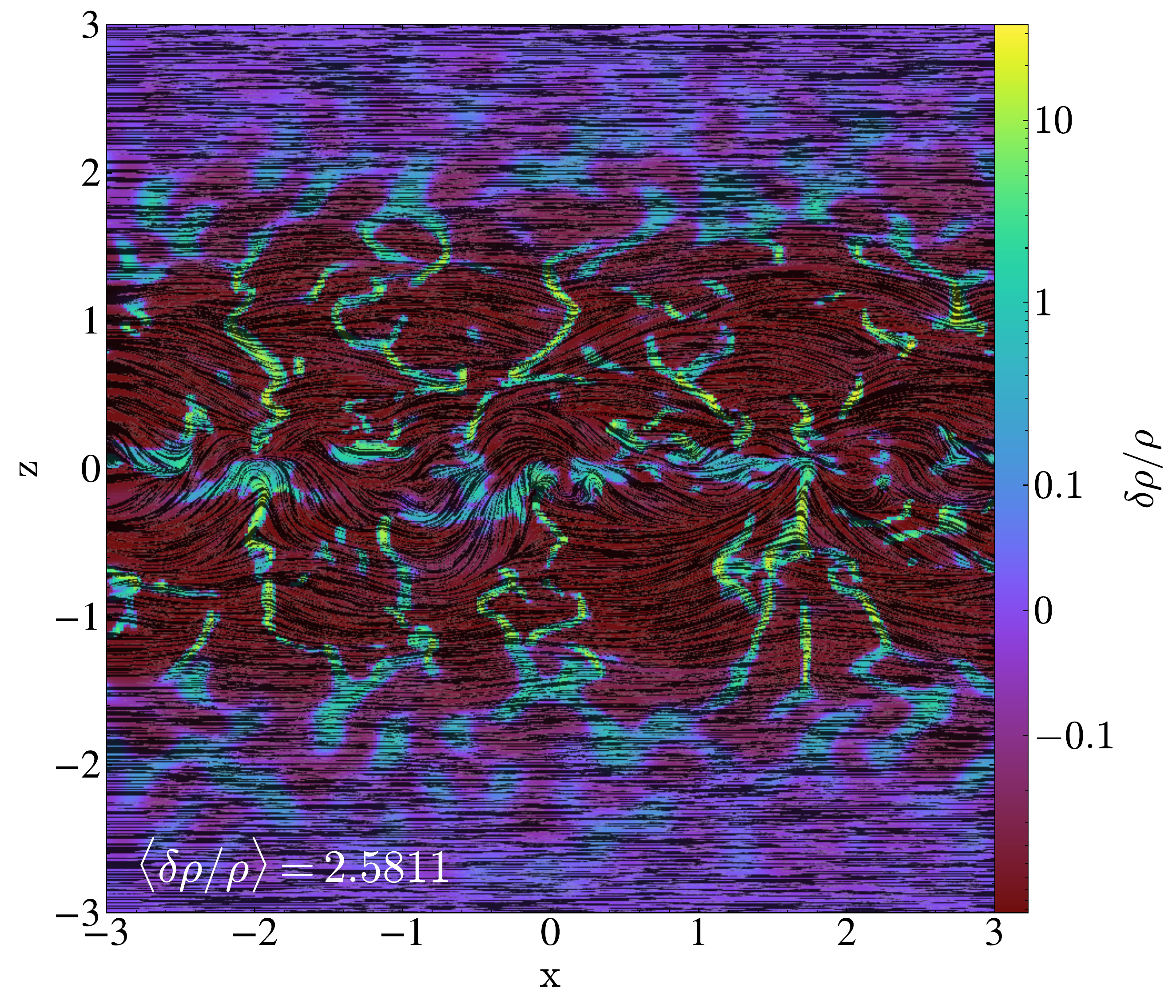}
      \caption{$\beta_0=3$, $t_\mathrm{cool}/t_\mathrm{ff} = 0.28$, cluster cooling curve}
      \label{fig:deltarho_beta3_cluster_short_cooling}
    \end{subfigure}
    \begin{subfigure}[b]{0.48\textwidth}
      \centering
      \includegraphics[width=\textwidth]{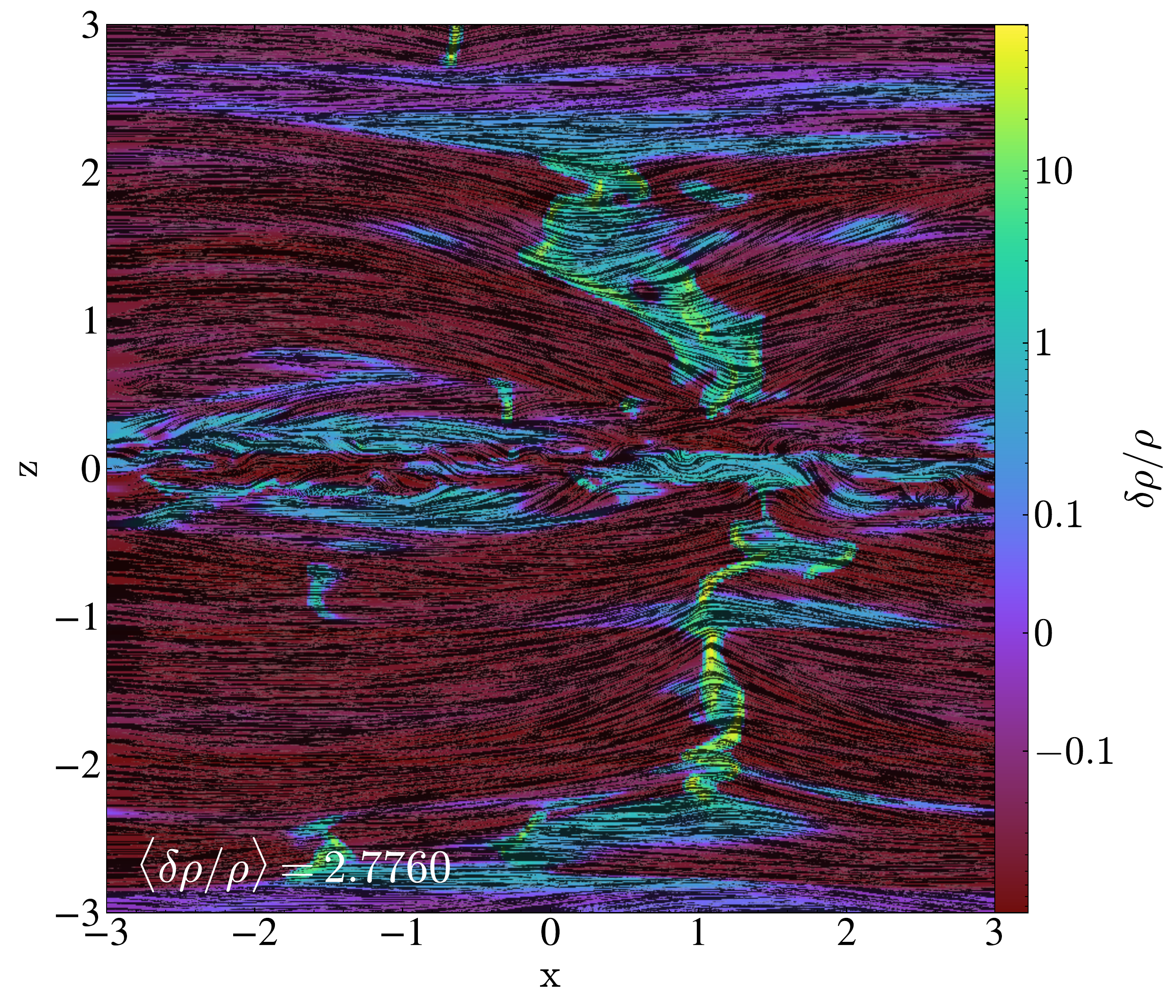}
      \caption{$\beta_0=3$, $t_\mathrm{cool}/t_\mathrm{ff} = 5.7$, galaxy cooling curve}
      \label{fig:deltarho_beta3_galaxy}
    \end{subfigure}
    \caption{Slice plots of the magnitude of density fluctuations with different initial $\beta_0$ values and cooling curves, superposed with magnetic field lines.}
    \label{fig:slice_deltarho}
  \end{figure*}

  In Fig. \ref{fig:deltarho_betainf_cluster} -- \ref{fig:deltarho_beta3_cluster}, we show slice plots of the magnitude of density fluctuations for simulations utilizing the cluster cooling curve, for $t_\mathrm{cool}/t_\mathrm{ff} = 5.7$ at $z=1$ (i.e., at one scale height) and initial $\beta_0=\infty\ (\text{hydro}), 278, 27, 3$ corresponding to RMS density fluctuations of $\langle \delta\rho/\rho\rangle_\mathrm{RMS} = 2.5 \times 10^{-2}, 9 \times10^{-2}, 0.15, 0.27$. Note that since the dynamic range of $\langle \delta\rho/\rho\rangle_\mathrm{RMS}$ is different for each plot, the colorbar scale is also different for each plot. Also superimposed are the magnetic field lines. While the mean $\beta$ in the entire box does not evolve significantly, it is clear that magnetic fields are amplified around overdense regions. The origin of this amplification can be deduced by evaluating the stretching and compressional terms in the flux freezing equation \citep{ji2016efficiency}:
  \begin{align}
  \frac{1}{2}\frac{d |\bm{B}|^2}{d t}
    = \overbrace{\bm{B} \cdot(\bm{B}\cdot \nabla)\bm{v} -\frac{1}{3} |\bm{B}|^2 \bm{\nabla}\cdot \bm{v}}^{\text{stretching}}
    \overbrace{-\frac{2}{3}|\bm{B}|^2 \bm{\nabla}\cdot \bm{v}}^{\text{compression}}.
  \end{align}

  For high $\beta$, the effects of stretching and compression are comparable, while in the low $\beta$ case, when magnetic tension is strong, amplification is weaker and mostly compressional. In particular, for weak initial fields, $\beta$ correlates strongly with overdensity and can drop by a factor of $\sim 10$ or more in regions within high overdensities; thus, magnetic fields can have a more significant dynamical effect than the mean or initial $\beta_0$ might indicate. Furthermore, as discussed in \S\ref{sect:interpretation} [see equations \eqref{eq:balance_z} and \eqref{eq:balance_perp}, and Fig. \ref{fig:tension}], what matters is not $\beta = P / P_\mathrm{B}$ but $\delta P/P_\mathrm{B} \sim O(1)$. The magnetic stresses should not be compared against pressure forces (which are nearly balanced by gravity), but with the perturbed pressure forces, with which they are comparable.

  In the hydrodynamic case, Fig. \ref{fig:deltarho_betainf_cluster}, density fluctuations grow at the largest available scales, and show clear horizontal stratification. This horizontal stratification is characteristic of low-frequency g-modes, where buoyancy forces inhibit vertical motions. By contrast, once magnetic fields are introduced, the morphology of density fluctuations changes completely. For very weak initial fields, for example with $\beta_0 = 278$ (Fig. \ref{fig:deltarho_beta278_cluster}), cold gas is vertically oriented and has much more small-scale structure. As the magnetic field strength increases ($\beta_0=27, 3$, Figs \ref{fig:deltarho_beta27_cluster} and \ref{fig:deltarho_beta3_cluster}), the vertical bias persists, but the density fluctuations appear on increasingly larger scales. Finally, if we re-run the simulation with strong magnetic fields $\beta_0=3$, but shorten the cooling time by a factor of $20$ ($t_\mathrm{cool}/t_\mathrm{ff}=0.28$, Fig. \ref{fig:deltarho_beta3_cluster_short_cooling}), then vertically oriented overdense filaments with much more small-scale structure appear. Later, we shall argue that these variations arise because the maximally unstable modes scale as $\sim v_\mathrm{A} t_\mathrm{cool}$; thus, weaker magnetic fields or shorter cooling times result in more structure at smaller scales. Eventually, as $\beta\sim1000$, the system converges to the hydrodynamic case, as we discuss further in \S\ref{sect:interpretation}.

\subsection{Impact of field orientation}
\label{sect:field_orient}

\begin{figure*}
    \begin{subfigure}[b]{0.48\textwidth}
      \centering
      \includegraphics[width=\textwidth]{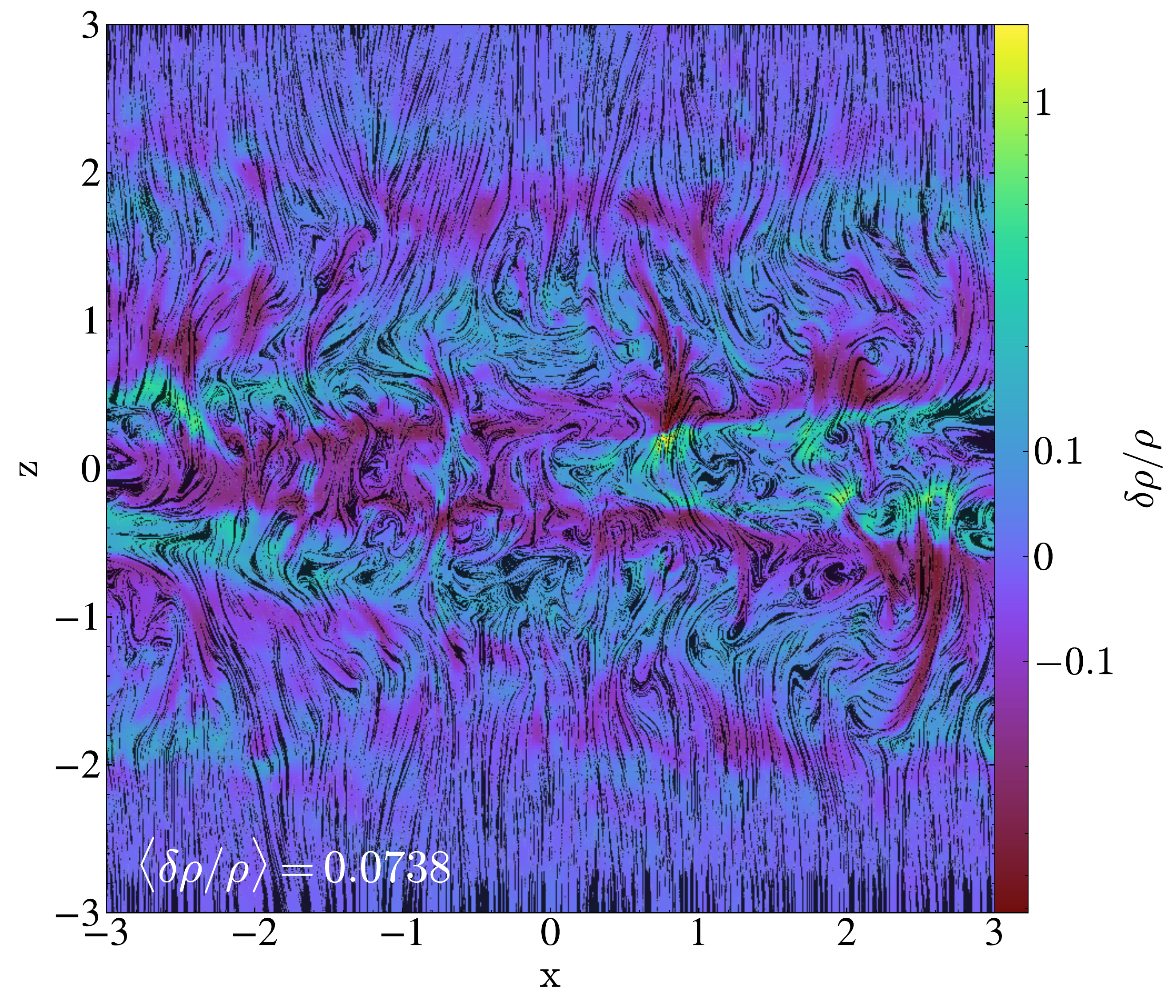}
      \caption{$\beta_0=278$, cluster cooling curve and initially vertical field}
      \label{fig:deltarho_beta278_cluster_by}
    \end{subfigure}
    \begin{subfigure}[b]{0.48\textwidth}
      \centering
      \includegraphics[width=\textwidth]{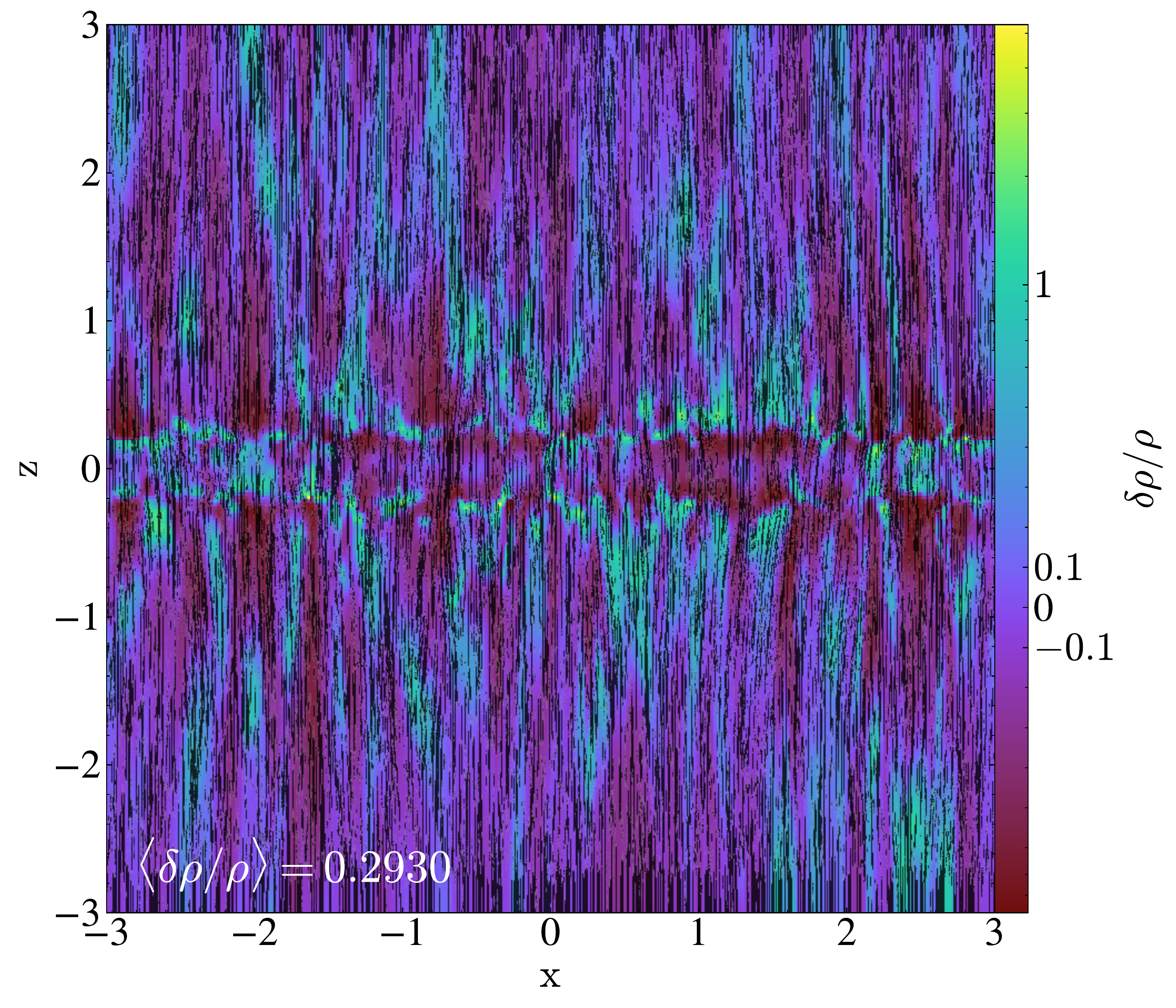}
      \caption{$\beta_0=3$, cluster cooling curve and initially vertical field}
      \label{fig:deltarho_beta3_cluster_by}
    \end{subfigure}
    \caption{Slice plots of the magnitude of density fluctuations with $t_\mathrm{cool}/t_\mathrm{ff} = 5.7$ and initially vertical magnetic fields, superposed with line integral convolution of magnetic vector field.}
    \label{fig:slice_deltarho_by}
  \end{figure*}

  Clearly, magnetic fields change the structure of g-modes seen in the hydrodynamic case. Since magnetic fields introduce anisotropic stresses, one might expect their impact to change with magnetic field orientation. For example, one might reasonably expect that in the horizontal field case, magnetic tension supports overdense gas and suppresses buoyancy-driven g-modes. This effect should vanish for vertically oriented fields. Indeed, when we run simulations with the same initial plasma $\beta_0$ and $t_\mathrm{cool}/t_\mathrm{ff} = 5.7$ but vertical fields, the morphology of density fluctuations appears completely different (Fig. \ref{fig:slice_deltarho_by}). For $\beta_0=278$ (Fig. \ref{fig:deltarho_beta278_cluster_by}, compare with Fig. \ref{fig:deltarho_beta278_cluster}), the density fluctuations appear larger scale and more planar. Note that the magnetic fields become more horizontal in overdense regions; there is some degree of magnetic tension support. By contrast, for $\beta_0=3$ (Fig. \ref{fig:deltarho_beta3_cluster_by}, compare with Fig. \ref{fig:deltarho_beta3_cluster}), density fluctuations appear at very small scales and manifest as thin vertical strips of alternating high and low density. We shall see that here, magnetic tension support is even more important (see Fig. \ref{fig:tension_by_lowbeta}).

  Incredibly, despite the strikingly different gas morphologies in simulations with initially horizontally and vertically aligned magnetic fields, the final RMS density fluctuations are remarkably independent of field orientation, and depend only on $\beta$ and $t_\mathrm{cool}/t_\mathrm{ff}$. In Fig. \ref{fig:rho_vs_time_cluster}, we indicate with stars the result of a few different simulations with vertical fields. We see that in every instance $\langle\delta\rho/\rho\rangle_\mathrm{RMS}$ closely corresponds to that derived from the equivalent horizontal field simulations with the same $\beta_0$ and $t_\mathrm{cool}/t_\mathrm{ff}$. This remarkable result suggests that the total amount of cold gas is insensitive to the (in general unknown) field orientation. This is extremely surprising, given how strongly the appearance of density fluctuations varies with field orientation; we return to this result in \S\ref{sect:interpretation}, below.

\subsection{Dependence on cooling curve}
\label{sect:cooling_curve}

\begin{figure}
    \begin{center}
      \includegraphics[width=0.5\textwidth]{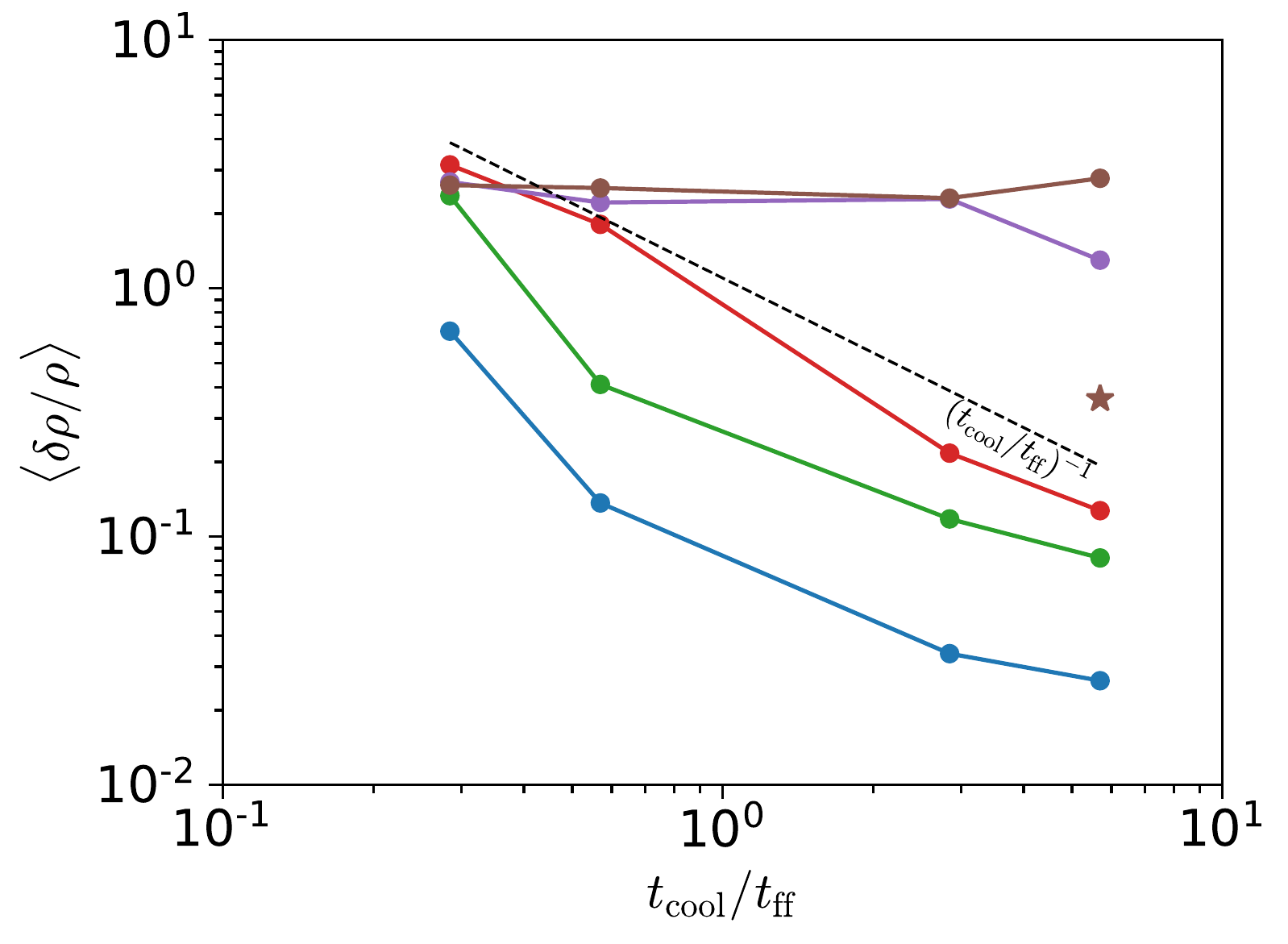}
      \caption{With galaxy cooling curve, $\delta \rho/\rho$ as a function of $t_\mathrm{cool}/t_\mathrm{ff}$ with different initial strengths of magnetic field characterized by the values of $\beta_0$: blue -- $\beta_0 = \infty$, green -- $\beta_0 = 278$, red -- $\beta_0=27$, purple -- $\beta_0=4$, brown -- $\beta_0=3$. The dot markers are simulations with initially horizontal field and star markers vertical field.}
      \label{fig:rho_vs_time_galaxy}
    \end{center}
  \end{figure}

  In Fig. \ref{fig:rho_vs_time_galaxy}, we show $\delta\rho/\rho$ as a function of $t_\mathrm{cool}/t_\mathrm{ff}$ for the galaxy-like cooling curve $\Lambda\propto T^{-1}$. For high $\beta$ ($\beta_0 = \infty, 278, 27$), the amplitude of the density fluctuations appears to be roughly similar to the cluster cooling curve case, with $\delta\rho/\rho \propto (t_\mathrm{cool}/t_\mathrm{ff})^{-1} \beta^{-1/2}$ independent of magnetic field orientation. However, at low $\beta$ ($\beta_0=3, 4$) and for horizontal fields, density fluctuations appear to saturate with a non-linear amplitude\footnote{The saturation amplitude $\delta\rho/\rho \sim 3$ is close to the maximum allowed $\delta\rho/\rho \sim \sqrt{\delta - 1} \sim 4.4$ as the mass fraction in cold gas with overdensity $\delta \sim T/T_\mathrm{floor} \sim 20$ approaches unity.} of $\delta\rho/\rho > 1$, \emph{independent} of $t_\mathrm{cool}/t_\mathrm{ff}$. At face value, this result implies when magnetic fields approach equipartition, gravity drops out of the problem and a multi-phase medium will always form, independent of $t_\mathrm{cool}/t_\mathrm{ff}$. Importantly, this only happens for horizontal fields: the vertical field results show the same $\delta\rho/\rho\propto (t_\mathrm{cool}/t_\mathrm{ff})^{-1}$ scaling at both high and low $\beta$, similar to the cluster cooling curve case. We show in \S\ref{sect:interpretation}, however, that closer examination reveals this to be actually a box size effect. In Fig. \ref{fig:rho_vs_time_galaxy}, we also show with a star the result of a simulation with the same effective resolution but double the vertical extent ($6 H \times 6H\times 12H$). The density fluctuation has fallen considerably, and is close to the $(t_\mathrm{cool}/t_\mathrm{ff})^{-1}$ scaling still obeyed by the cluster cooling curve. Examination of the slice plots for the smaller and larger boxes show that both only have one dominant filament, with $v_\mathrm{A} t_\mathrm{cool}>L_\mathrm{box}$ in the smaller box, suggesting that box size effects (\S\ref{sec:box_size}) might play a role. We will return to this point in \S\ref{sect:independence_cooling_curve}.

  The slice plots of the density field for $\beta_0=\infty, 278, 27$ for the galaxy cooling curve are similar to that for the cluster case and are not shown. More interesting are the slice plots for low $\beta$. It is interesting to contrast the galaxy cooling curve case, Fig. \ref{fig:deltarho_beta3_galaxy} ($\beta_0=3$, $t_\mathrm{cool}/t_\mathrm{ff}=5.7$, $\delta\rho/\rho = 2.7760$) with the cluster cooling curve case, Fig. \ref{fig:deltarho_beta3_cluster} ($\beta_0=3$, $t_\mathrm{cool}/t_\mathrm{ff}=5.7$, $\delta\rho/\rho = 0.2676$). Both of these slice plots slow the cold gas to be condensed in only $\sim 2$ vertically oriented structures in the entire simulation volume (one each above and below the midplane). This is in contrast to the $\beta_0 = 278, 27$ cases (Fig. \ref{fig:deltarho_beta278_cluster}, \ref{fig:deltarho_beta27_cluster}), where multiple such structures are visible. However, the cold gas in the galaxy cooling curve case reaches much higher peak overdensities, and thus has a much more pronounced filamentary structure. Finally, the contrast between Fig. \ref{fig:deltarho_beta3_galaxy} and Fig. \ref{fig:deltarho_beta3_cluster_short_cooling} ($\beta_0=3$, $t_\mathrm{cool}/t_\mathrm{ff} = 0.28$) is instructive. Although both have similar RMS density fluctuations and cold gas masses, they have very different topologies. Fig. \ref{fig:deltarho_beta3_cluster_short_cooling} shows much more small scale fragmentation and appears more similar to Fig. \ref{fig:deltarho_beta278_cluster}, albeit with a higher density contrast. We will discuss this in \S\ref{sect:interpretation} in light of the characteristic scale $v_\mathrm{A} t_\mathrm{cool}$.

  Except at low $\beta$ (discussed further in \S\ref{sect:independence_cooling_curve}), these results are consistent with a picture where the maximally unstable length scale (which we shall later identify as $\sim v_\mathrm{A} t_\mathrm{cool}$) depends only on $\beta$ and $t_\mathrm{cool}/t_\mathrm{ff}$, independent of the shape of the cooling curve.

\subsection{Convergence; 2D versus 3D}
\label{sect:conv}

\begin{figure*}
    \begin{subfigure}[t]{0.8\textwidth}
      \centering
      \includegraphics[width=\textwidth]{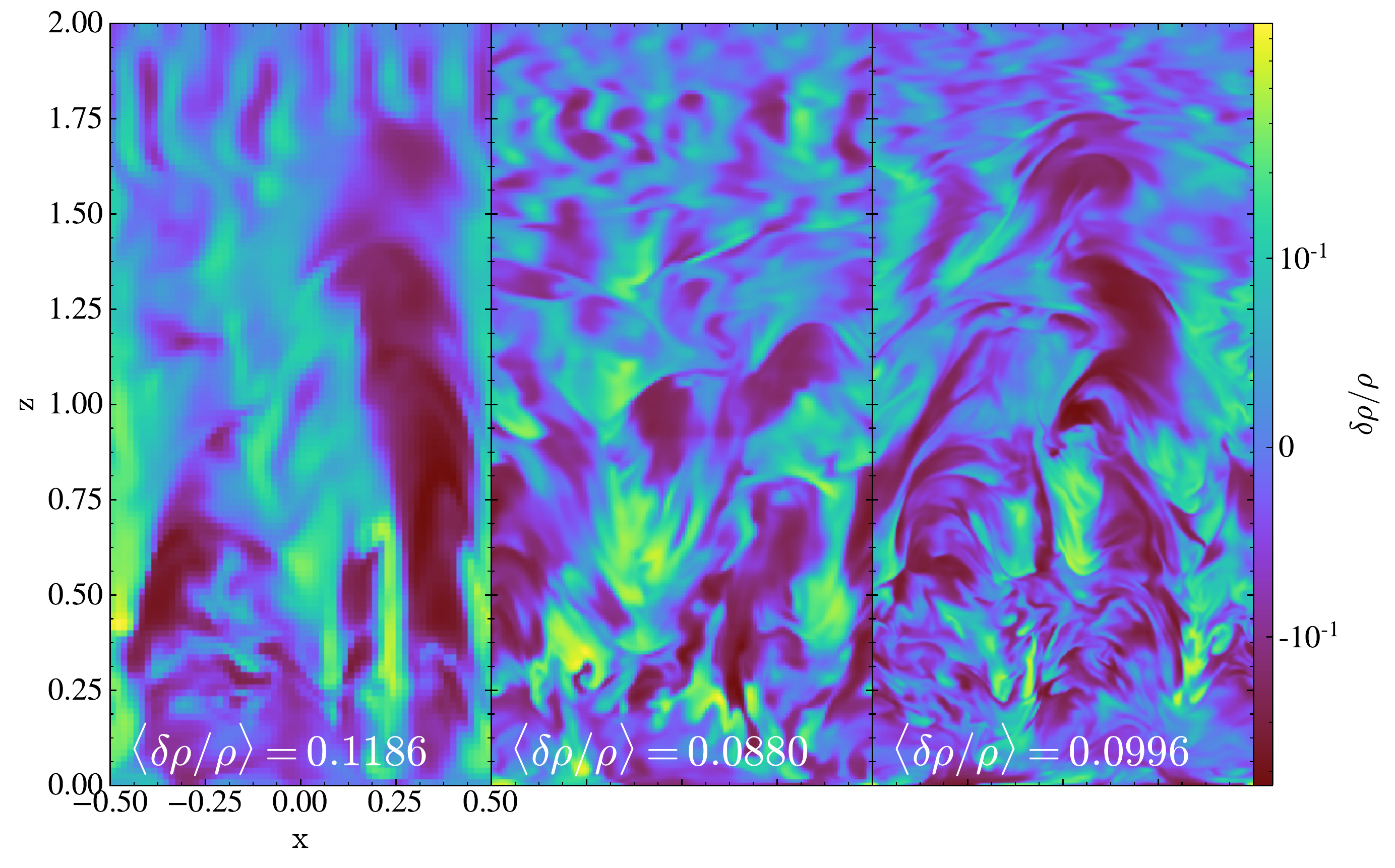}
      \caption{$\beta_0=278$}
      \label{fig:conv_b0.05}
    \end{subfigure}
    \begin{subfigure}[t]{0.8\textwidth}
      \centering
      \includegraphics[width=\textwidth]{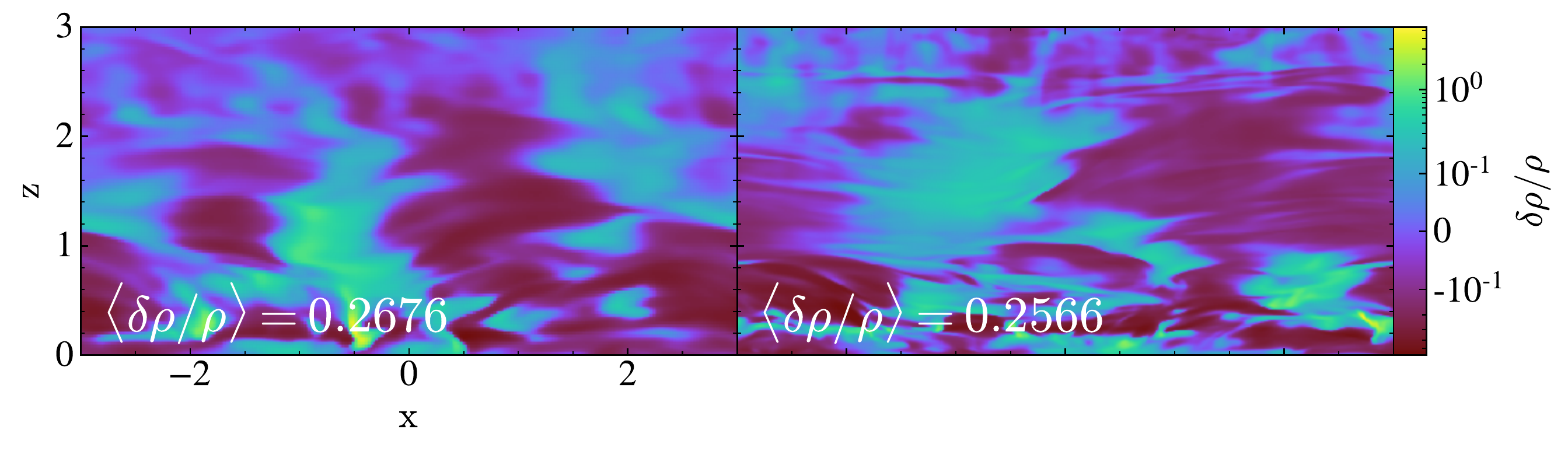}
      \caption{$\beta_0=3$}
      \label{fig:conv_b0.5}
    \end{subfigure}
    \caption{Convergence tests. Panel (\ref{fig:conv_b0.05}): domain size of $H\times H\times 2H$, and resolutions of $64\times64\times128$, $128\times128\times256$ and $256\times256\times512$ from left to right columns, at $t_\mathrm{cool}/t_\mathrm{ff}=5.7$. Panel (\ref{fig:conv_b0.5}): domain size of $6H\times 6H\times3H$, and resolutions of $256\times256\times128$ and $512\times512\times256$ from left to right columns, at $t_\mathrm{cool}/t_\mathrm{ff}=5.7$.}
    \label{fig:conv}
  \end{figure*}

  We show the results of convergence tests on a smaller box in Fig. \ref{fig:conv}. Our fiducial simulations are of size $6H$ (where $H$ is the scale height) and $256^3$, and thus have $\sim 42$ cells per scale height. It is also useful to phrase resolution in terms of the length scales:
  \begin{align}
    l_\mathrm{hydro}^\mathrm{cool} &\sim c_\mathrm{s} t_\mathrm{cool} \sim H \frac{t_\mathrm{cool}}{t_\mathrm{ff}} \\
    l_\mathrm{A}^\mathrm{cool} &\sim v_\mathrm{A} t_\mathrm{cool} \sim \frac{H}{\beta^{1/2}} \frac{t_\mathrm{cool}}{t_\mathrm{ff}},
  \end{align}
  where we have used $H\sim c_\mathrm{s} t_\mathrm{ff}$ and $\beta\sim(c_\mathrm{s}/v_\mathrm{A})^2$. We discuss the relevance of these length scales further in \S\ref{sect:scale}. While these length scales are generally well resolved, in some regions of parameter space where $t_\mathrm{cool}/t_\mathrm{ff}<1$, $\beta\gg1$, they are not. This applies to data points in the short cooling time limit at high $\beta$ in Fig. \ref{fig:rho_vs_time_cluster}, where our simulations may be less reliable.

  In Fig. \ref{fig:conv} we present the results of convergence studies for both high ($\beta_0=278$) and low ($\beta_0=3$) $\beta$ for a fixed value of $t_\mathrm{cool}/t_\mathrm{ff} = 5.7$ with horizontal fields and a cluster-like cooling curve. Tests in other regions of parameter space give similar results. In Fig. \ref{fig:conv_b0.05}, we show the results of simulations for $\beta_0=278$ with $\sim 64, 128, 256$ cells per scale height $H$ (or $\sim 13, 27, 54$ cells per $l_\mathrm{A}^\mathrm{cool}$). Note that unlike our fiducial simulations, here we only simulate the upper half ($z>0$) of the box. We see that while the detailed morphology of the slice plots changes (with more small scale detail apparent at higher resolution), the amplitude of density fluctuations is stable at the $\sim 10 - 20 \%$ level (and similar to that in our fiducial box, $\langle \delta\rho/\rho\rangle = 0.0899$). We shall argue that this is because density fluctuations peak at the scale $l_\mathrm{A}^\mathrm{cool}$, which is well resolved. Similarly, in Fig. \ref{fig:conv_b0.5}, we show the results of simulations for $\beta_0=3$, with $\sim 42, 85$ cells per $H$ (or $\sim 21, 42$ cells per $l_\mathrm{A}^\mathrm{cool}$), which also show converged values for $\langle\delta\rho/\rho\rangle$ despite differences in detailed morphology. Note the differing box size for these convergence tests ($H\times H \times 2 H$ in Fig. \ref{fig:conv_b0.05}, $6H\times 6H \times 3H$ in Fig. \ref{fig:conv_b0.5}). As is apparent in Fig. \ref{fig:conv_b0.5}, this change is necessary because of the much larger unstable modes in the low $\beta$ case. We will explore this further in the following section.

  We have also run several 2D simulations and compared them to our 3D results. In general, we find that 2D simulations produce higher density fluctuations by a factor of $2$ -- $3$. 2D simulations were already seen to produce somewhat higher density fluctuations in the hydrodynamic simulations of \citet{mccourt12}; the effect is more significant in MHD simulations. The reduced degrees of freedom constrains gas motions, particularly in the anisotropic MHD case, resulting in reduced damping of thermal instabilities. For this problem, it is important to use 3D simulations.

\subsection{Effect of box size}
\label{sec:box_size}

  \begin{figure*}
    \begin{subfigure}[t]{0.8\textwidth}
      \centering
      \includegraphics[width=\textwidth]{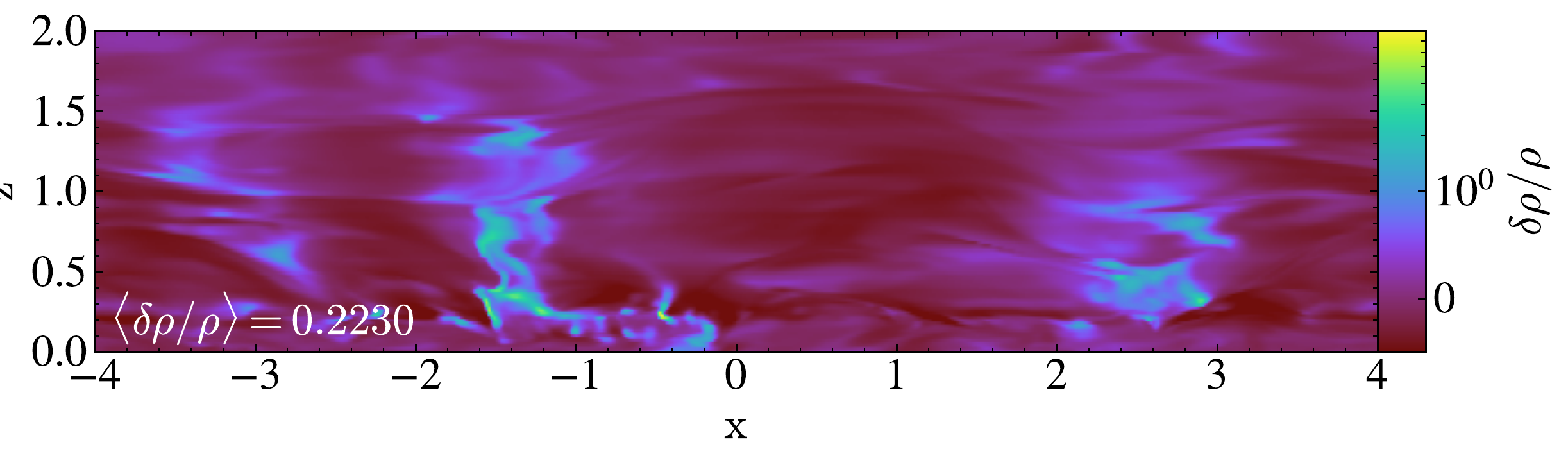}
      \caption{}
      \label{fig:over_cooling_longer}
    \end{subfigure} \\ 
    \begin{subfigure}[t]{0.6\textwidth}
      \centering
      \includegraphics[width=\textwidth]{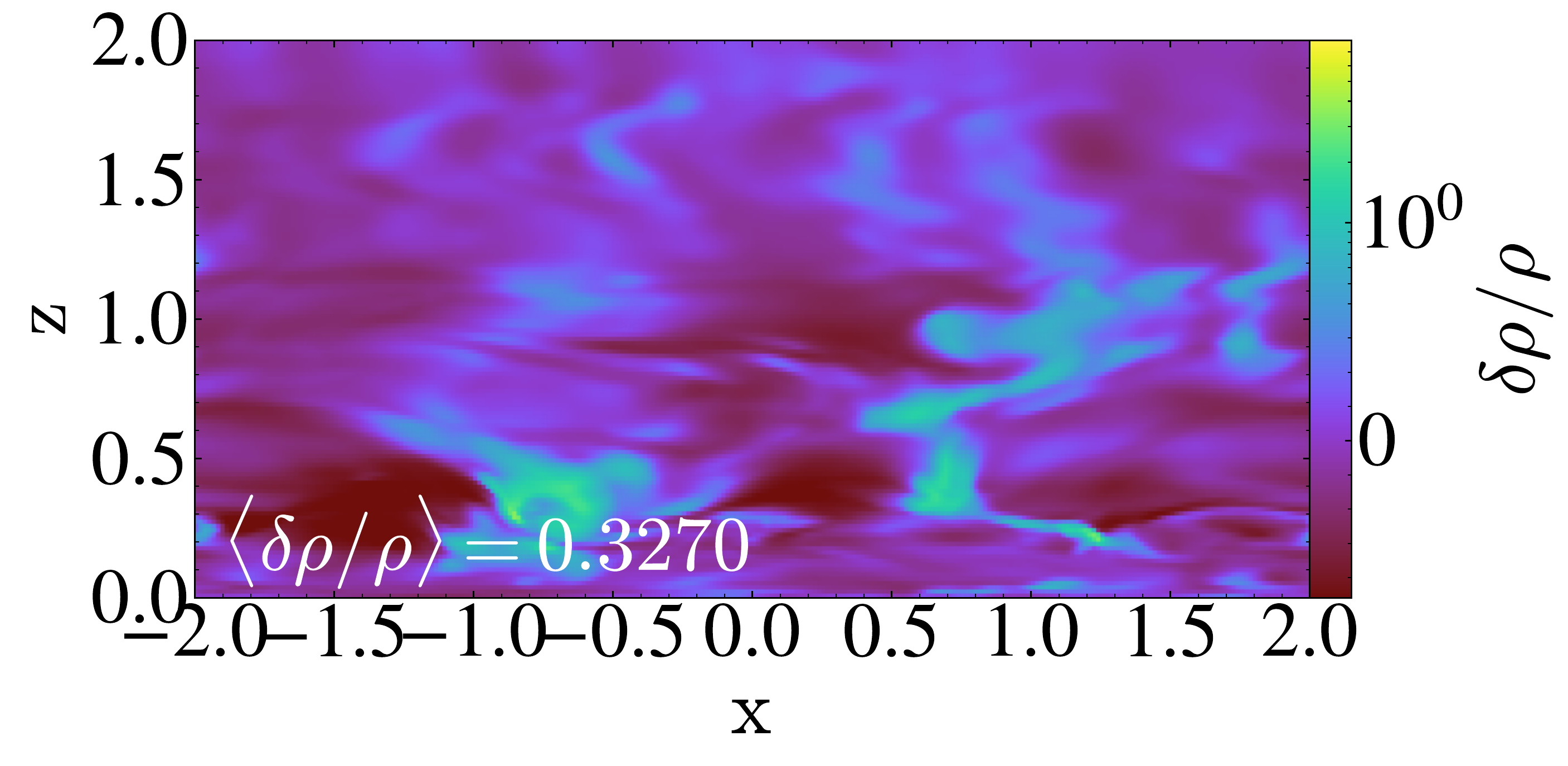}
      \caption{}
      \label{fig:over_cooling_long}
    \end{subfigure}
    \begin{subfigure}[t]{0.25\textwidth}
      \centering
      \includegraphics[width=\textwidth]{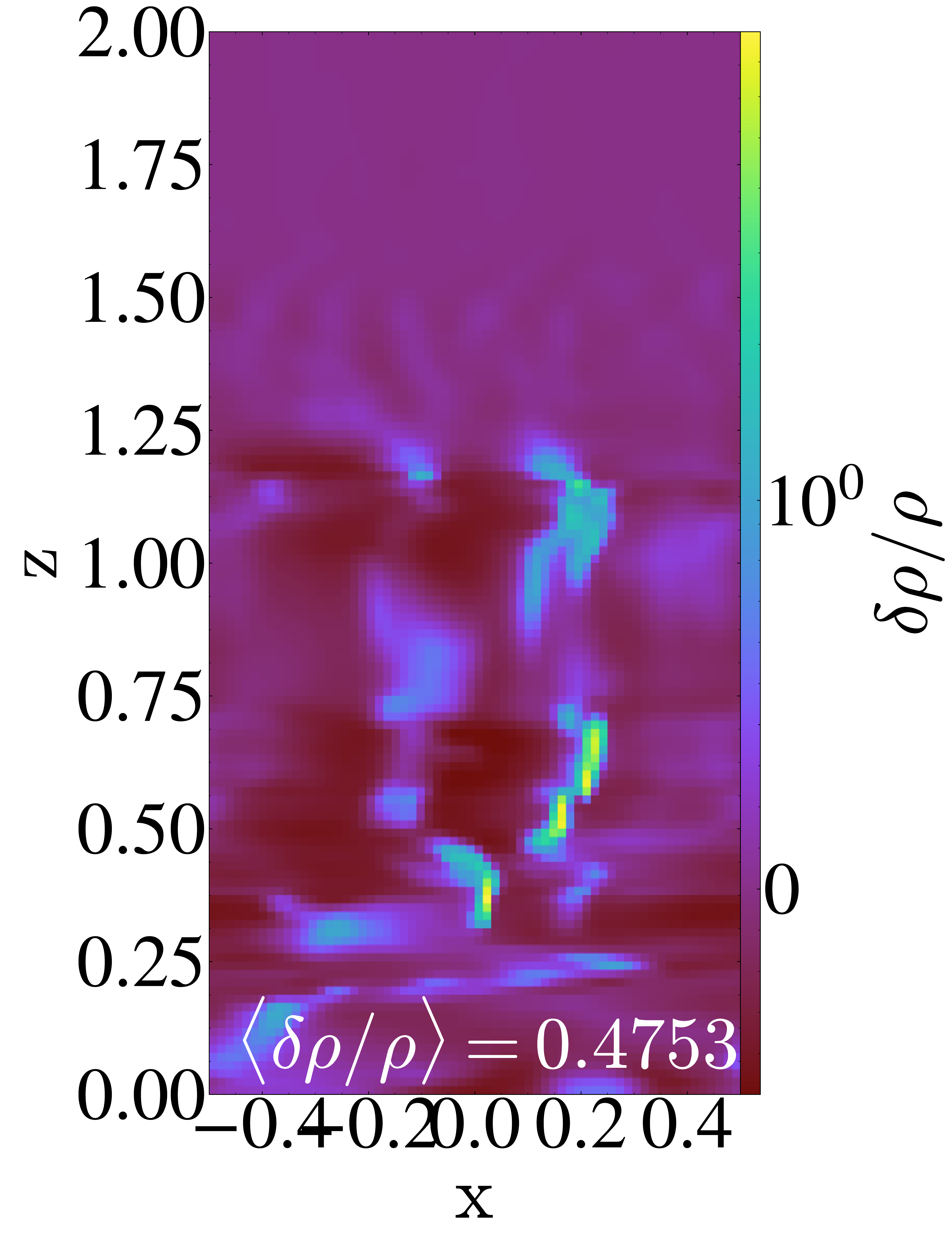}
      \caption{}
      \label{fig:over_cooling_short}
    \end{subfigure}
    \caption{Over-cooling effect: simulations of $t_\mathrm{cool}/t_\mathrm{ff} = 5.7$ and $\beta_0=3$ utilizing cluster cooling curve, with the same effective resolution of $64$ cells per $H$, but different box widths of $8\times4\times2$, $4\times4\times2$ and $1\times1\times2$.}
    \label{fig:over_cooling}
  \end{figure*}

  In Fig. \ref{fig:over_cooling}, we show the slice plots from simulations of different box size ($8\times 4\times2$, $4\times4\times2$, and $1\times1\times2$) but with the same effective resolution ($\sim 64$ cells per $H$, or $\sim 32$ cells per $l_\mathrm{A}^\mathrm{cool}$), for $t_\mathrm{cool}/t_\mathrm{ff} = 5.7$, $\beta_0=3$ and the cluster cooling curve. From the preceding section, this should produce converged RMS density fluctuations. Indeed, the $\delta\rho/\rho$ for the largest box is within $\sim 20\%$ of that of our fiducial simulation. However, for smaller boxes at fixed effective resolutions, $\delta\rho/\rho$ increases monotonically, and is double the fiducial value in the smallest box! This is despite the fact that the slice plots appear broadly similar, with $\sim 2$ large scale filaments per box. Later, we shall argue that this implies that $l_\mathrm{A}^\mathrm{cool}/L_\mathrm{box}$ is an important dimensionless parameter: we require $l_\mathrm{A}^\mathrm{cool} < L_\mathrm{box}$ for $\delta\rho/\rho$ to be independent of box size (for instance, in Fig. \ref{fig:over_cooling_short}, $L_\mathrm{box} < 0.5\ l_\mathrm{A}^\mathrm{cool}$, fails this criterion and thus produces larger overdensities).

\section{Interpretation}
\label{sect:interpretation}

  The preceding results show that magnetic fields impact both the amplitude and morphology of thermal instability in a stratified medium. The amplitude $\delta\rho/\rho\propto \beta^{-1/2}$ appears to be independent of both field orientation and cooling curve shape. However, the orientation of condensed filaments depends on both the amplitude and orientation of magnetic fields. Furthermore, the characteristic size of filaments changes with both plasma $\beta$ and $t_\mathrm{cool}/t_\mathrm{ff}$. We now provide a physical interpretation of these results.

\subsection{A characteristic scale}
\label{sect:scale}

  In order to understand the preceding results, it is useful to consider the dispersion relation for a magneto-gravity wave \citep{stein1974waves}:
  \begin{align}
    \omega^2 = \left(\frac{k_\perp}{k}\right)^2 \omega_\mathrm{BV}^2 + k_{\rm B}^2 v_\mathrm{A}^2,
    \label{eq:wave}
  \end{align}
  where $\omega_\mathrm{BV}^2 \sim g/H \sim t_\mathrm{ff}^{-2}$ is the Brunt-V\"ais\"al\"a frequency, $k_\perp$ indicates transverse wave numbers perpendicular to the direction of gravity, and $k_{\rm B} \equiv \bm{k}\cdot \hat{\bm b}_{0}$ is the wavenumber parallel to the background field direction. The two terms on the right-hand side represent g-mode and Alfv\'en-wave oscillations, where the restoring forces are buoyancy forces and magnetic tension, respectively. At first blush, it would appear from equation \eqref{eq:wave} that interchange modes (with variations perpendicular to the background field), should be unaffected by magnetic fields, and could play an important role. We shall soon see that our results are remarkable insensitive to the orientation of magnetic fields with respect to gravity. We can consider thermal instability to be a process which is driven at a frequency of $\omega_\mathrm{drive} \sim t_\mathrm{cool}^{-1}$, and is modified by the two processes represented by the right hand side of Eq. \eqref{eq:wave}, g-modes driven by buoyancy forces and Alfv\'en waves driven by magnetic tension forces. In the absence of non-linear effects, this simply leads to exponentially growing overstable modes. Below, we shall argue that non-linear damping has effective frequencies which scales as the linear oscillation frequencies.

  We can estimate the magnitude of density fluctuations in the hydrodynamic case as follows \citep{mccourt12}. The linearized momentum equation:
  \begin{align}
    \rho \frac{dv}{dt} = \delta \rho g
    \label{eq:mom}
  \end{align}
  gives a characteristic velocity of
  \begin{align}
    v \sim \frac{\delta\rho}{\rho} g t_\mathrm{buoy} \sim \frac{\delta\rho}{\rho} \frac{H}{t_\mathrm{ff}},
    \label{eq:ch_velocity}
  \end{align}
  where we have used $t_\mathrm{buoy} \sim t_\mathrm{ff}$ for the buoyant oscillation period and $g\sim H / t_\mathrm{ff}^2$, where $H$ is the pressure scale-height. The condition $t_\mathrm{drive} \sim t_\mathrm{damp}$ yields the fastest growing perturbations if we maximize $t_\mathrm{damp} \sim L/v$, setting $L\sim H$. Setting $t_\mathrm{drive}\sim t_\mathrm{cool}$, we thus obtain:
  \begin{align}
    \frac{\delta\rho}{\rho} \sim A_1 \left(\frac{t_\mathrm{cool}}{t_\mathrm{ff}}\right)^{-1}
    \label{eq:delrho_hydro}
  \end{align}
  in good agreement with our numerical results (Eq. \eqref{eq:delrho_time}; Fig. \ref{fig:rho_vs_beta}), where in planar geometry the dimensionless coefficient $A_1 \sim 0.1$.

  Eq. \eqref{eq:mom} is clearly modified in MHD case, when magnetic stresses have to be considered. For instance, with an initially horizontal field, magnetic tension opposes gravity, and an overdense fluid element can remain static. This suppression of buoyant oscillations can be shown formally in a linear stability analysis, resulting in cooling rates essentially identical to Field-type condensation modes in an unstratified medium \citep{loewenstein90, balbus91}. However, this effect is clearly scale-dependent. Magnetic stresses can only be significant during thermal instability if the Alfv\'en crossing time across a perturbation is less than the cooling time $l/ v_\mathrm{A} < t_\mathrm{cool}$, which implies that only length scales:
  \begin{align}
    l < l_\mathrm{A}^\mathrm{cool} \equiv v_\mathrm{A} t_\mathrm{cool}
    \label{eq:lengthscale}
  \end{align}
  are destabilized by magnetic fields. Note that in the hydrodynamic case, the largest scale modes are the most unstable, whereas in the MHD case, it is the small scale modes which grow fastest. This accounts for the distinct change in morphology between hydrodynamic and MHD simulations, even at high $\beta$. Contrast Fig. \ref{fig:deltarho_betainf_cluster} (hydro) and Fig. \ref{fig:deltarho_beta278_cluster} ($\beta_0\sim 300$); the latter shows much more small scale structure.

  The length scale in Eq. \eqref{eq:lengthscale} differs from that identified by \citet{loewenstein90} and  \cite{balbus91} in their linear analysis, who find that modes on scales smaller than $l_\mathrm{A} \sim H/\sqrt{\beta}$ are destabilized. The length scale $l_\mathrm{A}^\mathrm{cool} \sim v_\mathrm{A} t_\mathrm{cool}$ emerges from balancing the first and third terms of Eq. \eqref{eq:wave}, assuming $\omega\sim\omega_\mathrm{drive}\sim t_\mathrm{cool}^{-1}$, corresponding to a critical length scale where magnetic stresses affect thermal instability. By contrast, the length scale $l_\mathrm{A} \sim H / \sqrt{\beta}$ emerges from balancing the second and third terms of Eq. \eqref{eq:wave}, indicating when magnetic tension overwhelms buoyancy forces. Physically we require that magnetic tension beats both cooling and buoyancy forces. These two differ by a factor of $t_\mathrm{cool}/ t_\mathrm{ff}$:
  \begin{align}
    l_\mathrm{A}^\mathrm{cool} \sim v_\mathrm{A} t_\mathrm{cool} \sim c_\mathrm{s} \beta^{-1/2} t_\mathrm{cool} \sim H \beta^{-1/2} \frac{t_\mathrm{cool}}{t_\mathrm{ff}} \sim l_\mathrm{A} \frac{t_\mathrm{cool}}{t_\mathrm{ff}}.
  \end{align}
  Our simulations show that the length scale $l_\mathrm{A}^\mathrm{cool}$ (Eq. \eqref{eq:lengthscale}) is the relevant one, not $l_\mathrm{A}$ (contrast Fig. \ref{fig:deltarho_beta3_cluster} and Fig. \ref{fig:deltarho_beta3_cluster_short_cooling}, where $l_\mathrm{A}$ is the same in both simulations since $\beta$ is the same, but $l_\mathrm{A}^\mathrm{cool}$ is smaller by a factor of $\sim 20$ in Fig. \ref{fig:deltarho_beta3_cluster_short_cooling}, due to the smaller value of $t_\mathrm{cool}/t_\mathrm{ff}$. The latter clearly shows much more small scale structure).

  From these considerations, we can understand how magnetic stresses change the amplitude of density fluctuations. Tension forces balance gravity when
  \begin{align}
    \delta \rho g \sim \frac{B^2}{\lambda_\mathrm{B}},
    \label{eq:grav_tension}
  \end{align}
  where the magnetic tension force is $(\bm{B} \cdot \nabla){\bm B} = \hat{n} B^{2}/\lambda_{\rm B}$, where the unit vector $\hat{n}$ points to the center of the radius of curvature, and $\lambda_{\rm B}$ is the radius of curvature. Note that this is a non-linear criteria, where $\delta B \sim B$.    
  Setting $\lambda_\mathrm{B} \sim l_\mathrm{A}^\mathrm{cool}$, the maximal scale which can be influenced by magnetic tension during thermal instability, we obtain:
  \begin{align}
    \frac{\delta \rho}{\rho} \sim A_2 \left(\frac{t_\mathrm{cool}}{t_\mathrm{ff}}\right)^{-1} \beta^{-1/2}
    \label{eq:delrho_mhd}
  \end{align}
  which is consistent with our numerical results (Eq. \eqref{eq:delrho_time}; Fig. \ref{fig:rho_vs_time_cluster} and Fig. \ref{fig:rho_vs_beta}) which show $\delta\rho/\rho \propto \beta^{-1/2}$.

  Comparing Eq. \eqref{eq:delrho_hydro} and Eq. \eqref{eq:delrho_mhd}, we see that the MHD density fluctuations approaches the hydrodynamic case when
  \begin{align}
    \beta\sim \left(\frac{A_2}{A_1}\right)^2 \sim 900
  \end{align}
  where we have taken $A_1$, $A_2$ from Eq. \eqref{eq:delrho_time}. From Fig. \ref{fig:rho_vs_time_cluster}, we see that this is indeed the case. Thus, magnetic stresses start to become important even when the magnetic energy is highly subdominant.

  The characteristic scale $l_\mathrm{A}^\mathrm{cool}\sim v_\mathrm{A} t^\mathrm{cool}$ also explains why large boxes are needed for converged results in the low $\beta$, long cooling time limit. If the box is smaller than $l_\mathrm{A}^\mathrm{cool}$, then all modes within the box are destabilized by magnetic fields (for instance, Fig. \ref{fig:over_cooling_short}, where $L_\mathrm{box} < 0.5 l_\mathrm{A}^\mathrm{cool}$); gas motions are mostly strongly quenched for these small scale modes. Since damping of gas motions is artificially boosted, density fluctuations are larger for small boxes. Only when the largest unstable mode $l_\mathrm{A}^\mathrm{cool}$ is contained within the box do the RMS density fluctuations become independent of box size.

  It is useful to directly examine evidence from the simulations that magnetic tension from horizontal magnetic fields suppresses buoyant oscillations. In Fig. \ref{fig:part}, we present tracer particle plots for particles which manage to eventually cool for a range of initial heights above mid-plane. These particles are selected so that the maximum entropy decrease between $t = 0$ -- $17\ t_\mathrm{cool}$ is maximized. We show both the height  above mid-plane ($z$-coordinate) as a function of time, and spatial trajectories projected on to the $x$-$z$ plane (shifted to $x=0$ at $t=0$).

  In the hydrodynamic case (Fig. \ref{fig:part_betainf_cluster}), as expected, cooling particles execute over-stable buoyant oscillations. Only the particles at low height $|z|<1$, where $t_\mathrm{cool}/t_\mathrm{ff}$ is low, become overdense and sink. Note that entropy changes take place over many multiples of the cooling time, since the departure from thermal equilibrium is small. As expected from the stable background entropy stratification (and as is also clear from Fig. \ref{fig:deltarho_betainf_cluster}), particles execute mostly planar motions.\footnote{In a stably stratified medium, $v_2/v_1 \sim (v /\omega_\mathrm{BV} H)^2 \ll 1$ for weak turbulence (e.g., see section 2 of \citealt{ruszkowski10}) . Note that strong turbulence can overwhelm buoyant restoring forces and thereby promote thermal instability (Voit 2017, in preparation).}

  \begin{figure*}
    \begin{subfigure}[b]{\textwidth}
      \centering
      \includegraphics[width=0.43\textwidth]{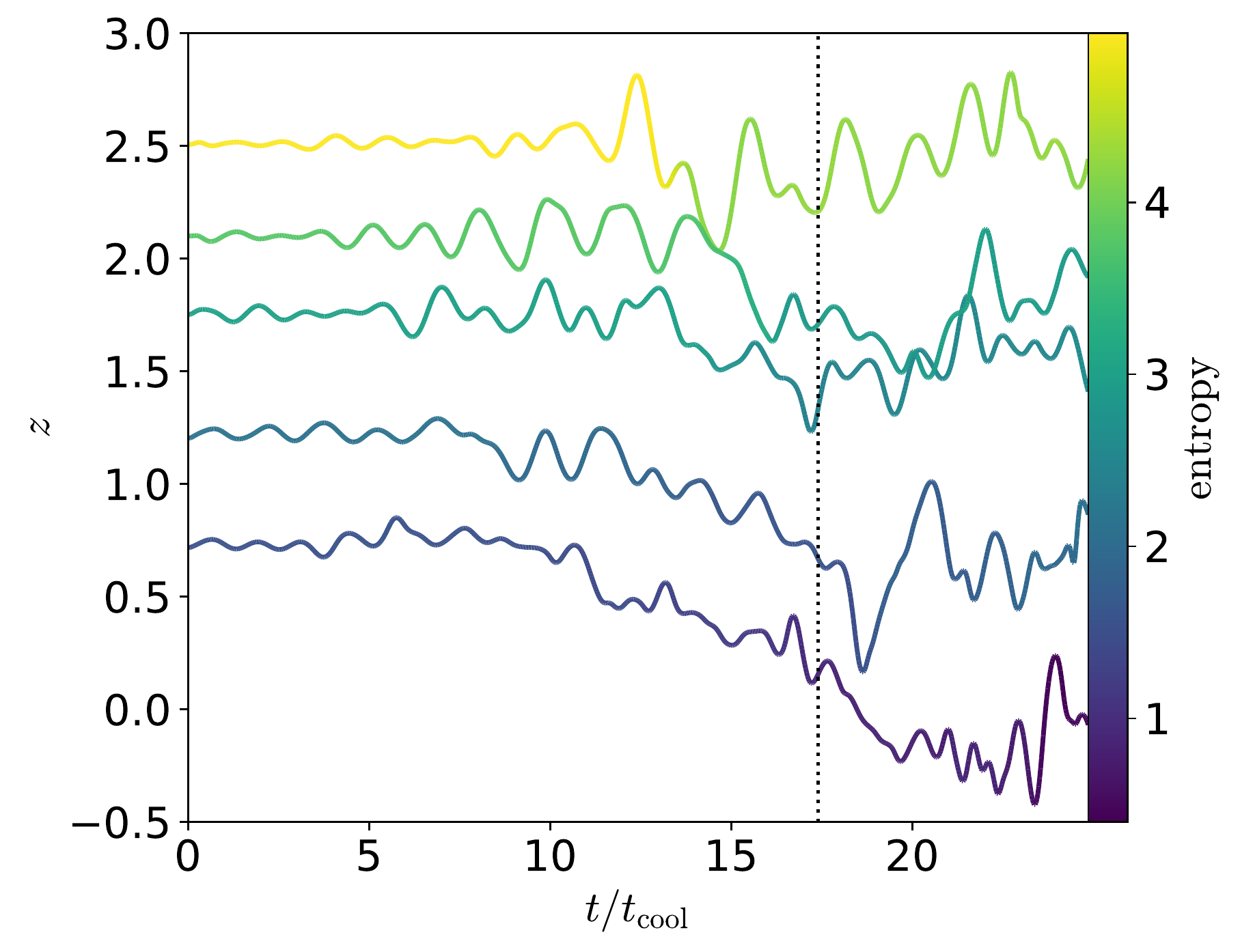}
      \includegraphics[width=0.54\textwidth]{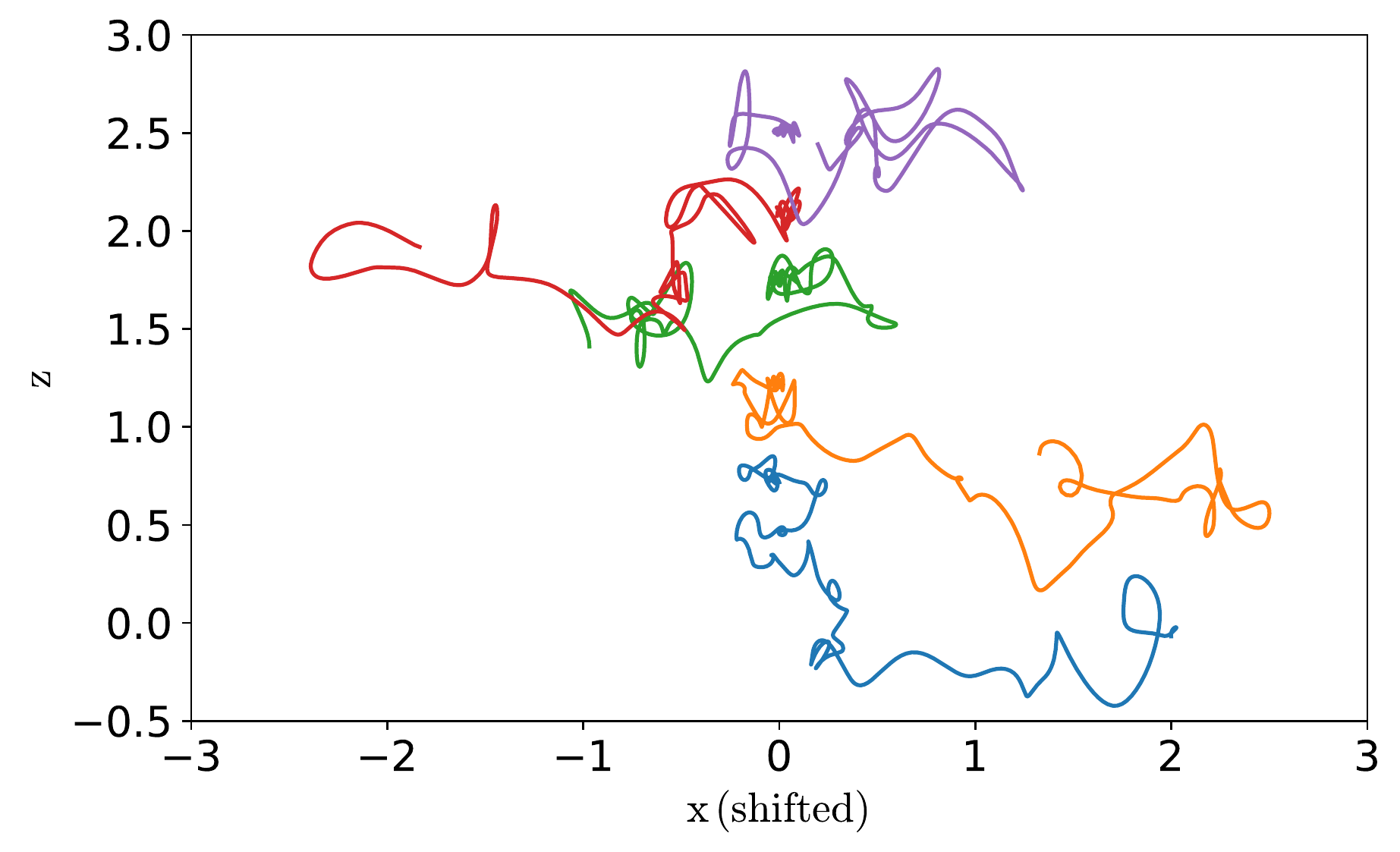}
      \caption{$\beta_0=\infty$}
      \label{fig:part_betainf_cluster}
    \end{subfigure}
    \begin{subfigure}[b]{\textwidth}
      \centering
      \includegraphics[width=0.44\textwidth]{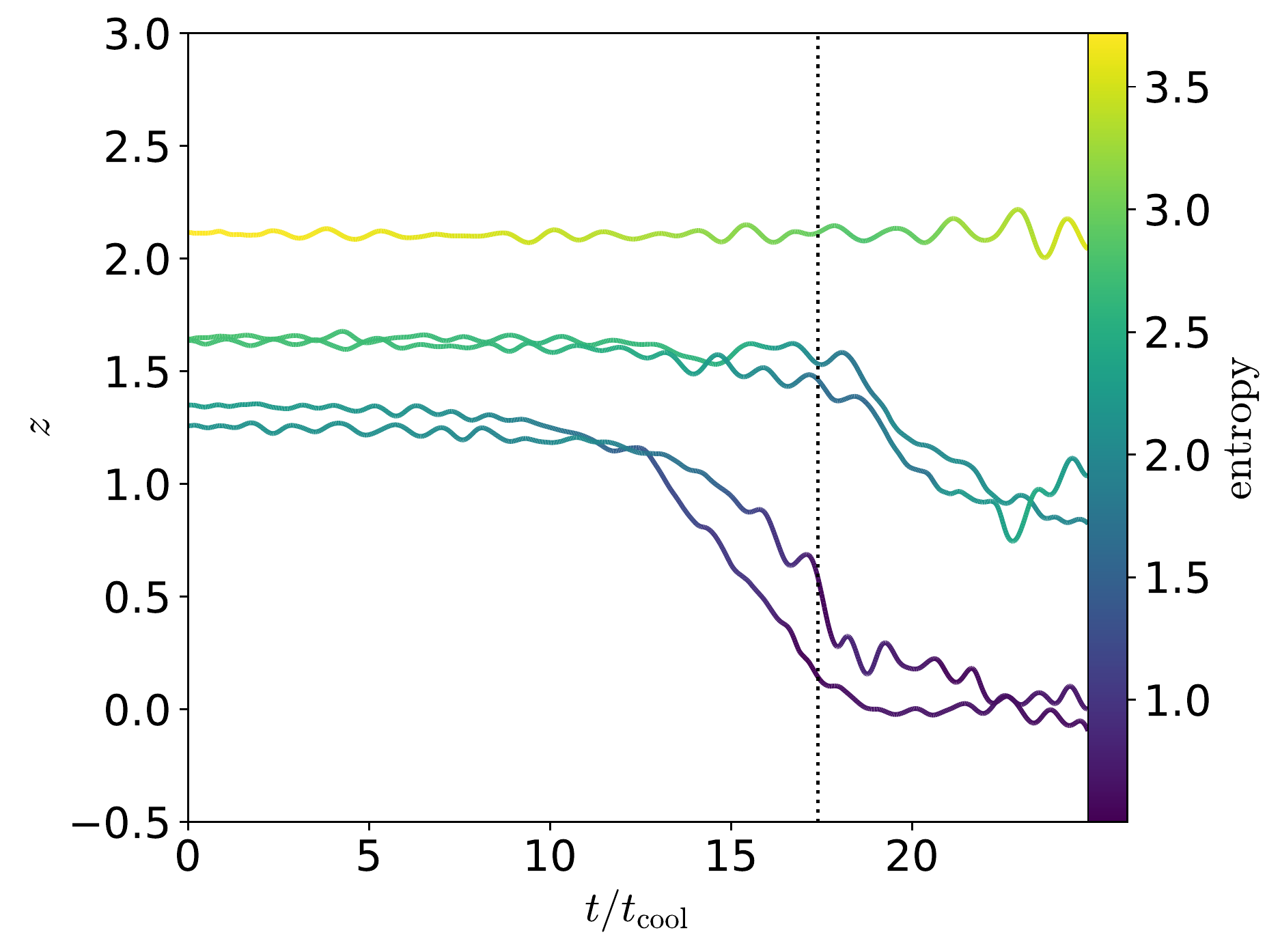}
      \includegraphics[width=0.53\textwidth]{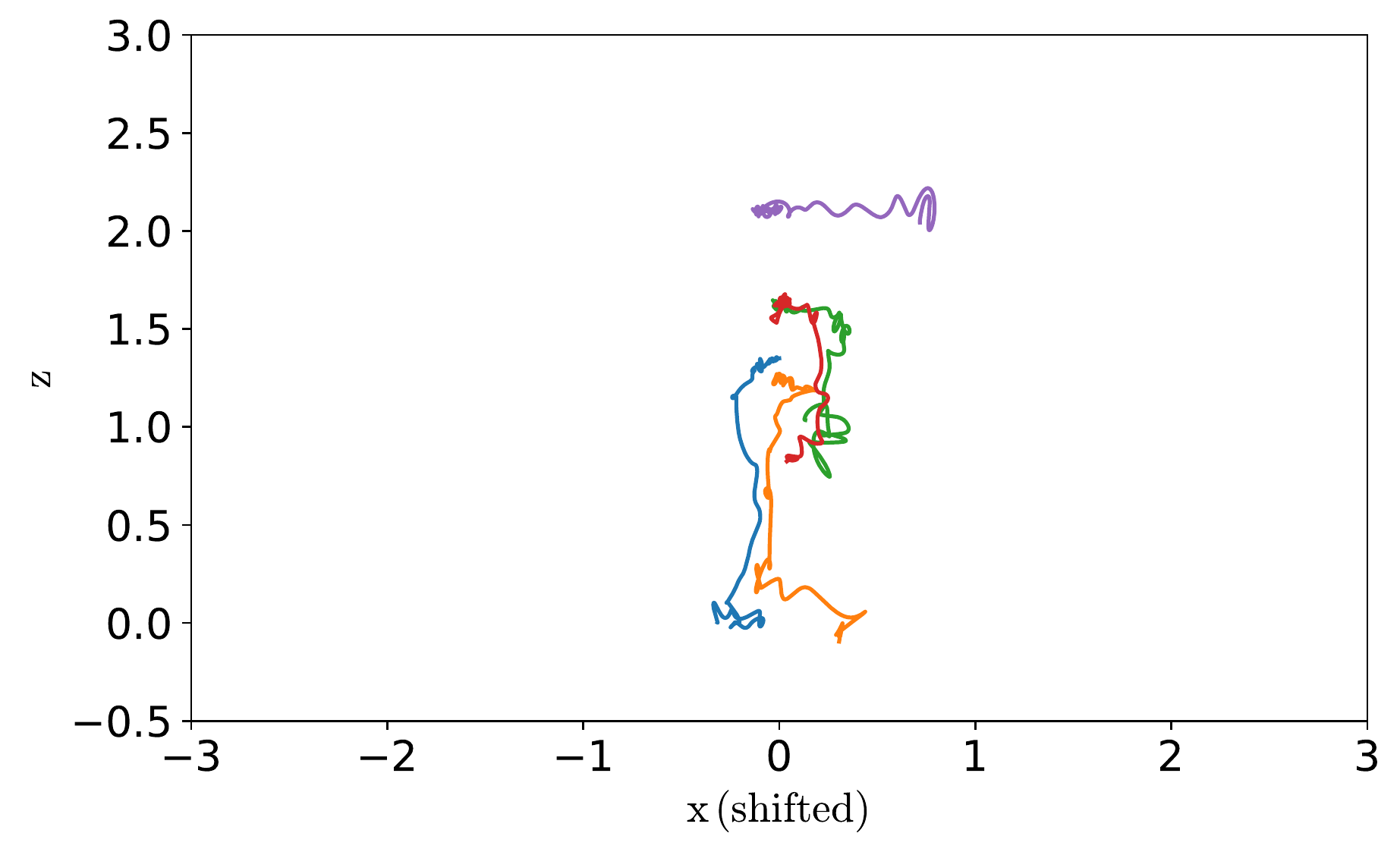}
      \caption{$\beta_0=278$, horizontal magnetic fields}
      \label{fig:part_beta278_cluster}
    \end{subfigure}
    \begin{subfigure}[b]{\textwidth}
      \centering
      \includegraphics[width=0.43\textwidth]{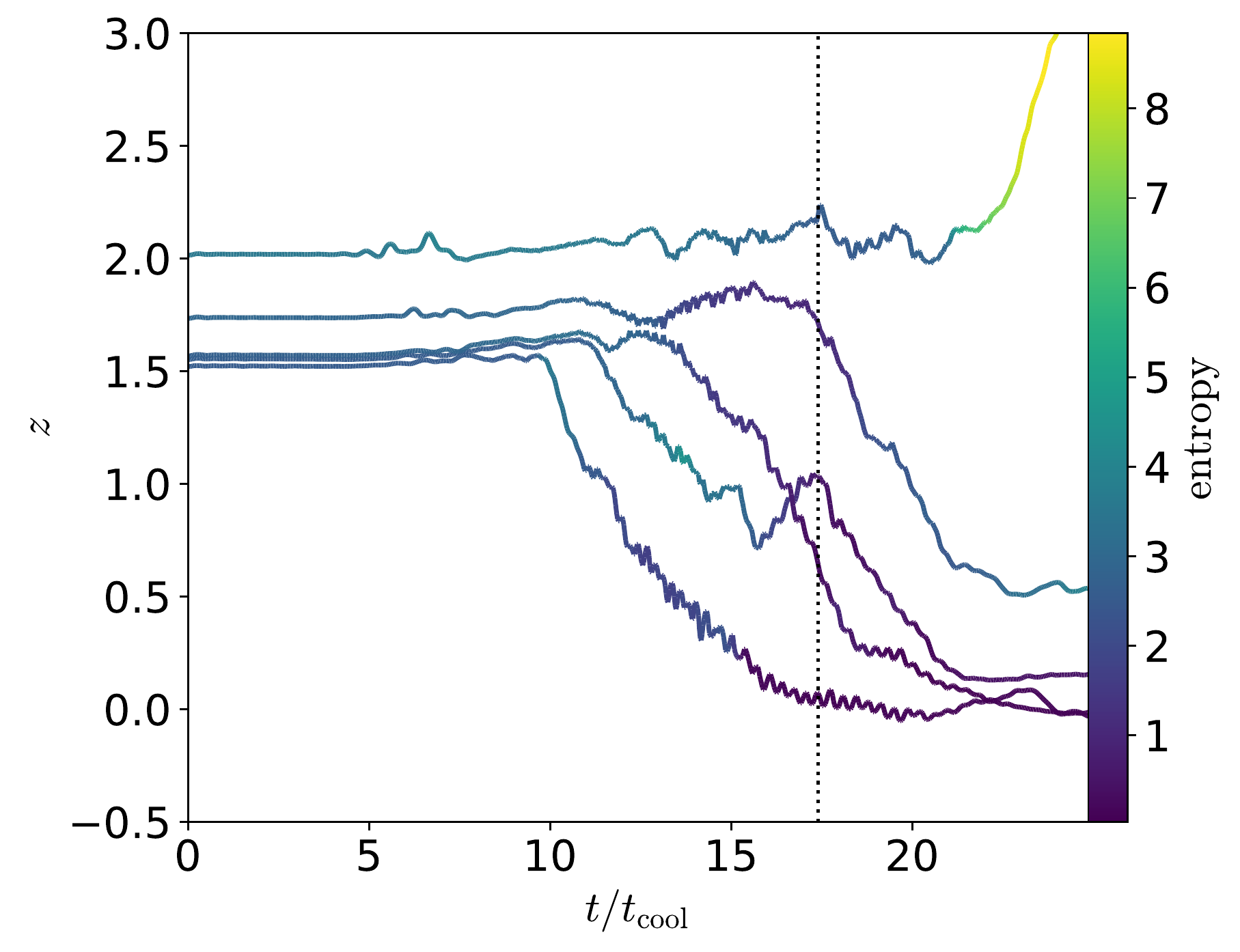}
      \includegraphics[width=0.54\textwidth]{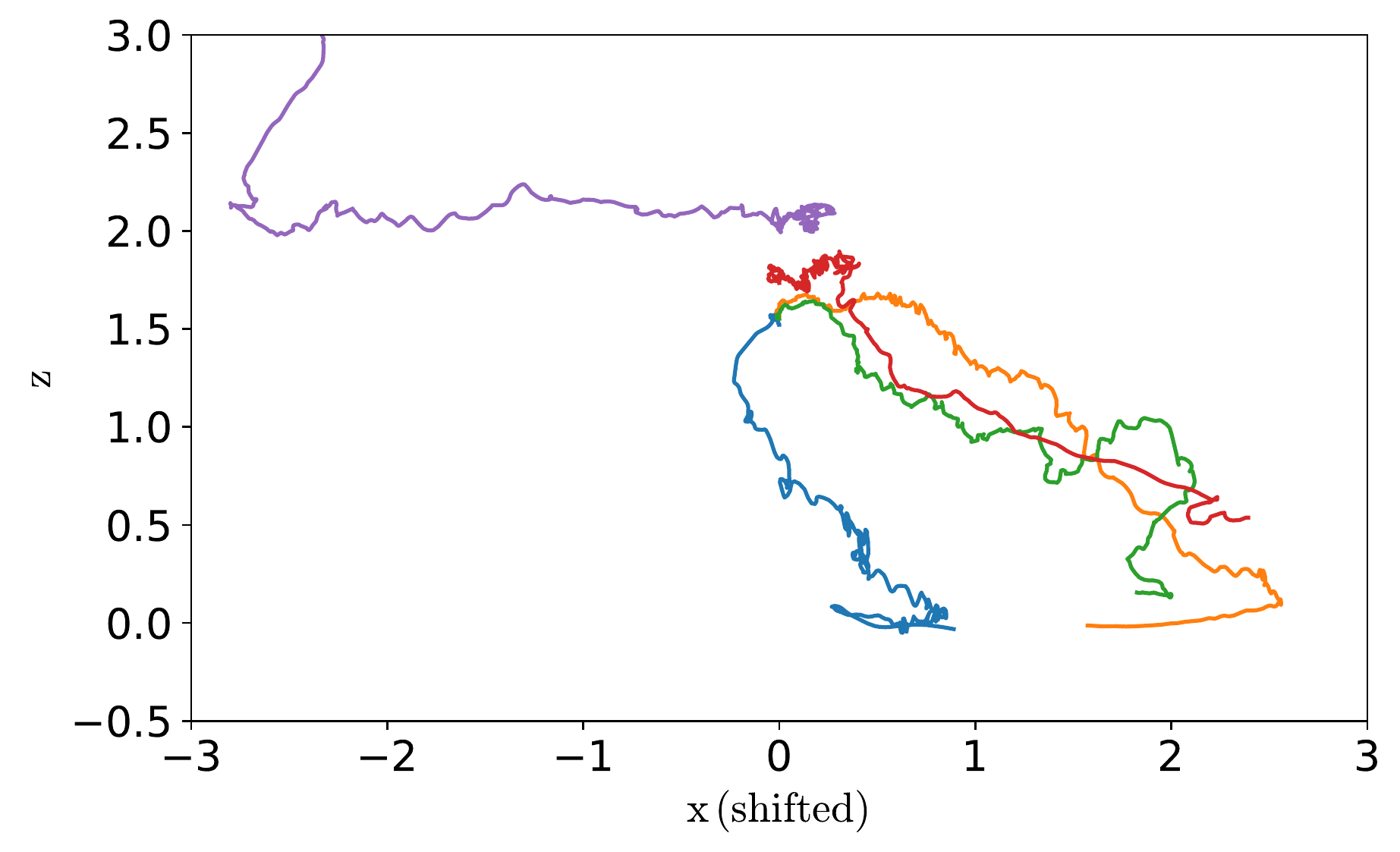}
      \caption{$\beta_0=3$, horizontal magnetic fields}
      \label{fig:part_beta3_cluster}
    \end{subfigure}
  \end{figure*}
  \begin{figure*}
    \ContinuedFloat
    \begin{subfigure}[b]{\textwidth}
      \centering
      \includegraphics[width=0.44\textwidth]{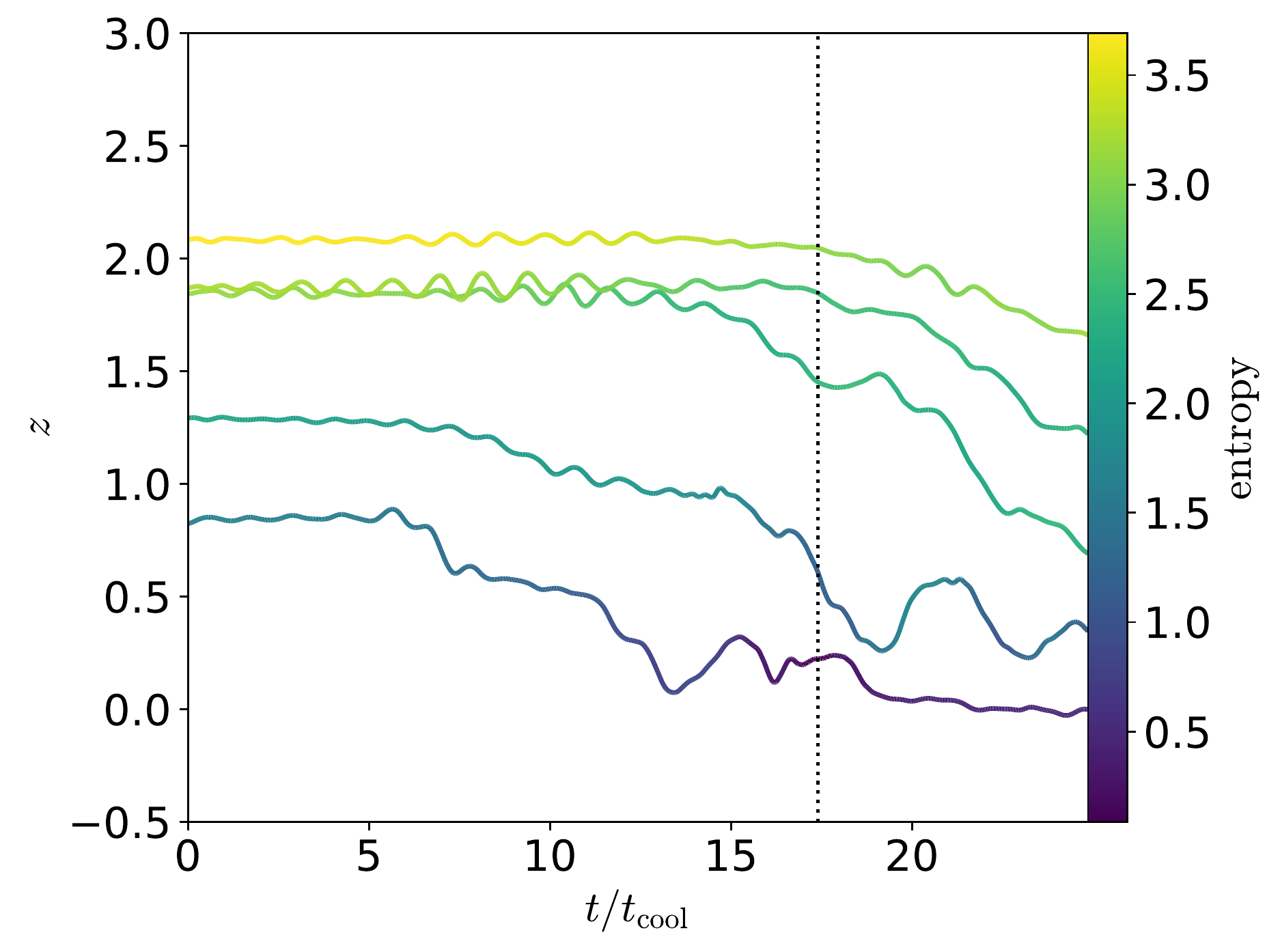}
      \includegraphics[width=0.53\textwidth]{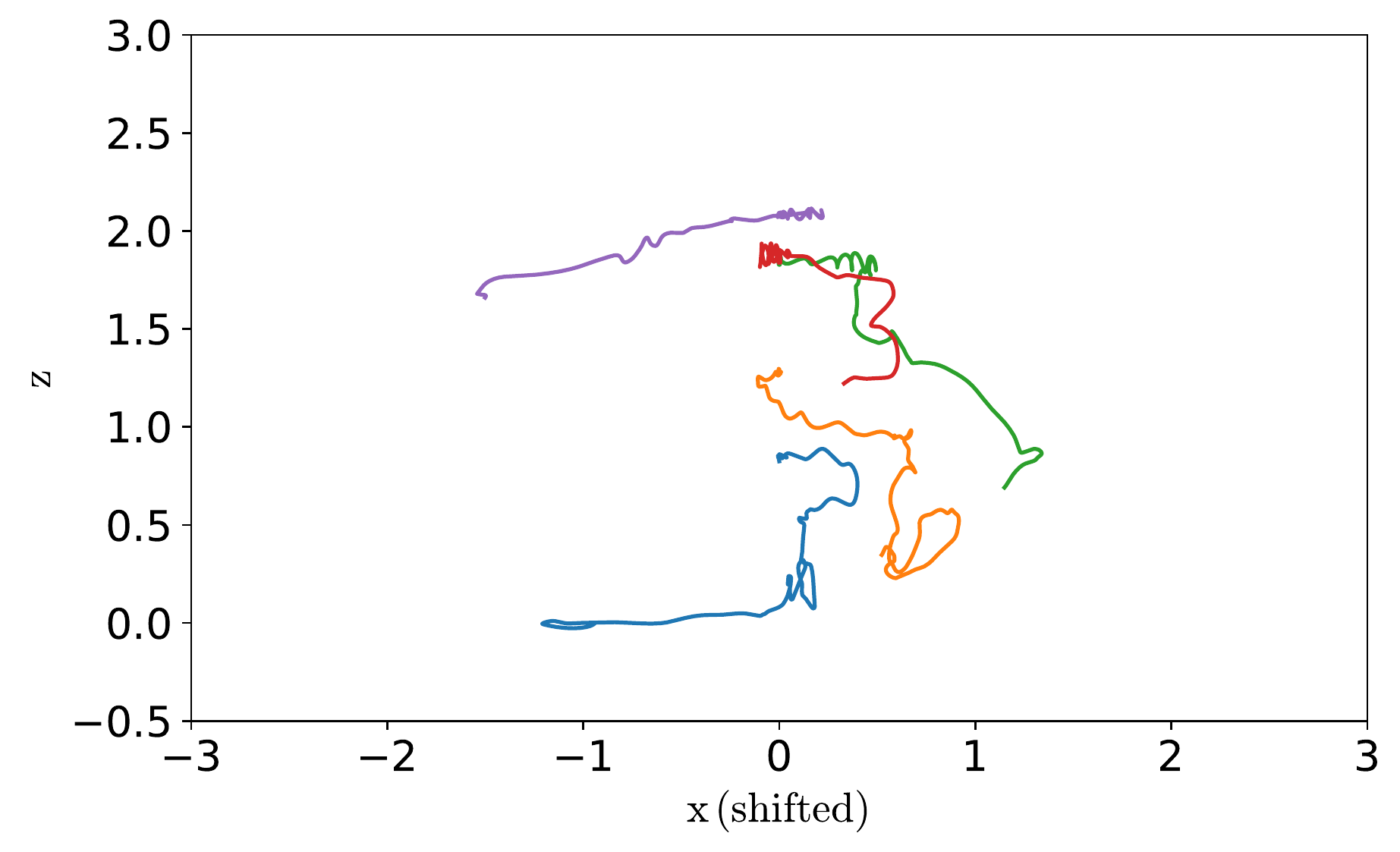}
      \caption{$\beta_0=278$, vertical magnetic fields}
      \label{fig:part_beta278_by_cluster_by}
    \end{subfigure}
    \begin{subfigure}[b]{\textwidth}
      \centering
      \includegraphics[width=0.43\textwidth]{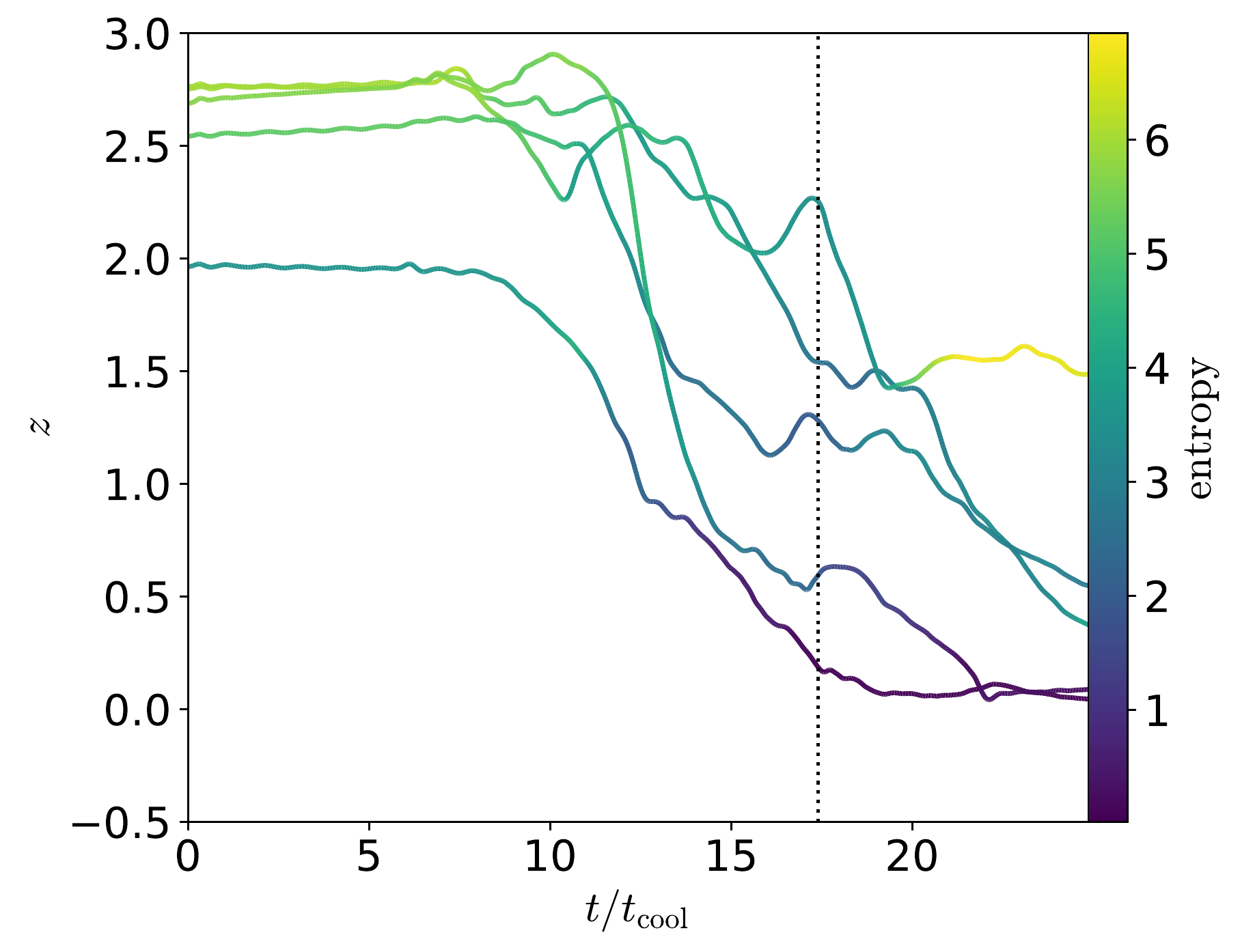}
      \includegraphics[width=0.54\textwidth]{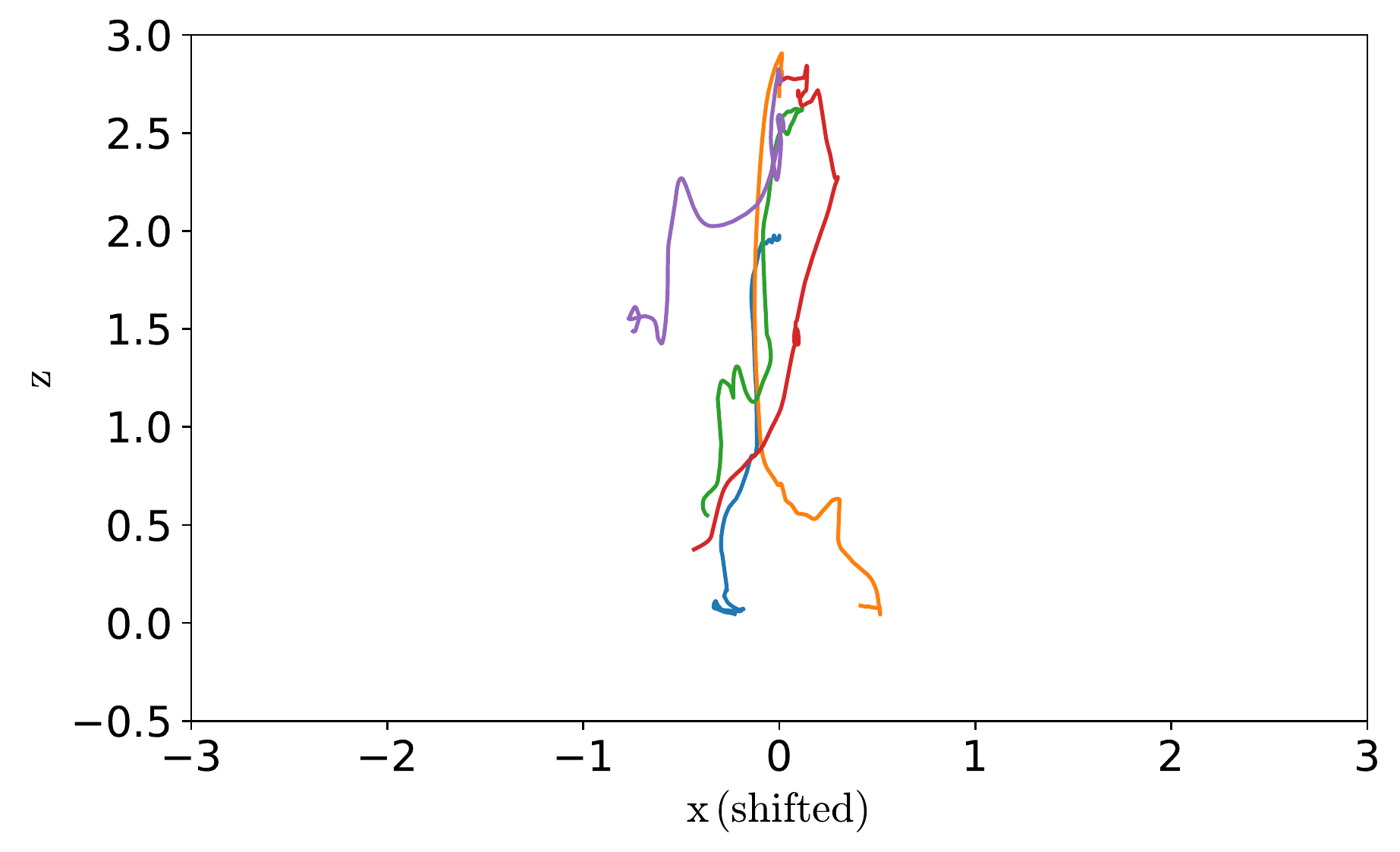}
      \caption{$\beta_0=3$, vertical magnetic fields}
      \label{fig:part_beta3_by_cluster_by}
    \end{subfigure}
    \caption{Selected particle temporal trajectories colored by entropy (left-hand panel) and spatial trajectories (right-hand panel, projected to $x$-$z$ plane and horizontally shifted to $x=0$ for starting point). The criterion for particle selection is the maximum entropy decrease from $t=0$ to $t\sim17\ t_\mathrm{cool}$ (dotted vertical line) at different heights, which helps to trace the pathway of cold gas formation. All the cases presented above are at long cooling time ($t_\mathrm{cool}/t_\mathrm{ff} = 5.7$).}
    \label{fig:part}
  \end{figure*}

  Once a horizontal magnetic field is introduced, we can see that buoyant oscillations are strongly suppressed (Fig. \ref{fig:part_beta278_cluster} and Fig. \ref{fig:part_beta3_cluster}). Instead, particles cool and lose entropy at a fixed height until they reach sufficient over-densities that they cannot be supported by magnetic tension. At that point, they fall. As the magnetic field strength is increased, buoyant oscillations are more strongly damped, and particles cool earlier at higher $z$, where $t_\mathrm{cool}/t_\mathrm{ff}$ is larger (compare left-hand panels of Fig. \ref{fig:part_beta278_cluster} and Fig. \ref{fig:part_beta3_cluster}). When the field is weak, the characteristic length scale $l_\mathrm{A}^\mathrm{cool}\sim v_\mathrm{A} t_\mathrm{cool}$ is small. Particles only travel a small horizontal distance in a cooling time before they cannot be supported by the magnetic field and fall (right-hand panel of Fig. \ref{fig:part_beta278_cluster}). This accounts for the small scale, vertically-oriented morphology of overdense filaments (Fig. \ref{fig:deltarho_beta278_cluster}). By contrast, when fields are strong ($\beta_0=3$), $l_\mathrm{A}^\mathrm{cool}$ is large, and particles can travel a considerable horizontal distance as they cool (right-hand panel of Fig. \ref{fig:part_beta3_cluster}), resulting in filaments which are not necessarily vertically oriented (Fig. \ref{fig:deltarho_beta3_cluster}).

  We can also directly verify the importance of magnetic tension in supporting overdense filaments, by examining terms in the linearized, perturbed Euler equation. In Fig. \ref{fig:tension_bx}, we compare the gravitational force on an overdense filament with the perturbed pressure gradient $\partial_z \delta P_\mathrm{tot}$ (where $P_\mathrm{tot} = P_\mathrm{g} + P_\mathrm{B}$) and the perturbed magnetic tension forces $(\bm{B}\cdot\nabla) \bm{B}_z / 4\pi$, for the $\beta_0 = 278$, horizontal field case. We note that the magnetic tension term corresponds closely in morphology to the $\delta \rho g$ term, providing most $\sim 70\%$ of the support against gravity. On the other hand, pressure gradients contribute a non-negligible amount ($\sim 30\%$) and can reach comparable peak amplitudes, although the spatial correspondence is less clear, and their time averaged influence is less important.

  \begin{figure*}
    \begin{subfigure}[b]{\textwidth}
      \centering
      \includegraphics[width=\textwidth]{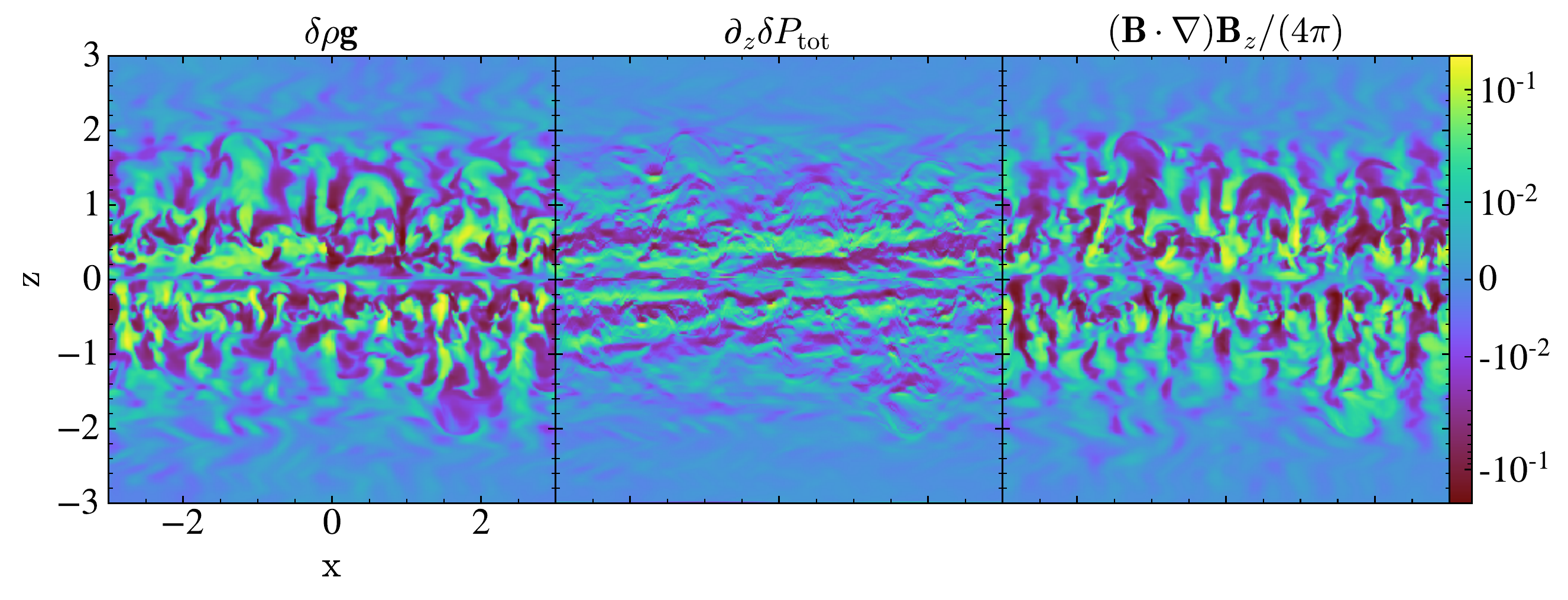}
      \caption{$\beta_0=278$, horizontal field}
      \label{fig:tension_bx}
    \end{subfigure}
    \begin{subfigure}[b]{\textwidth}
      \centering
      \includegraphics[width=\textwidth]{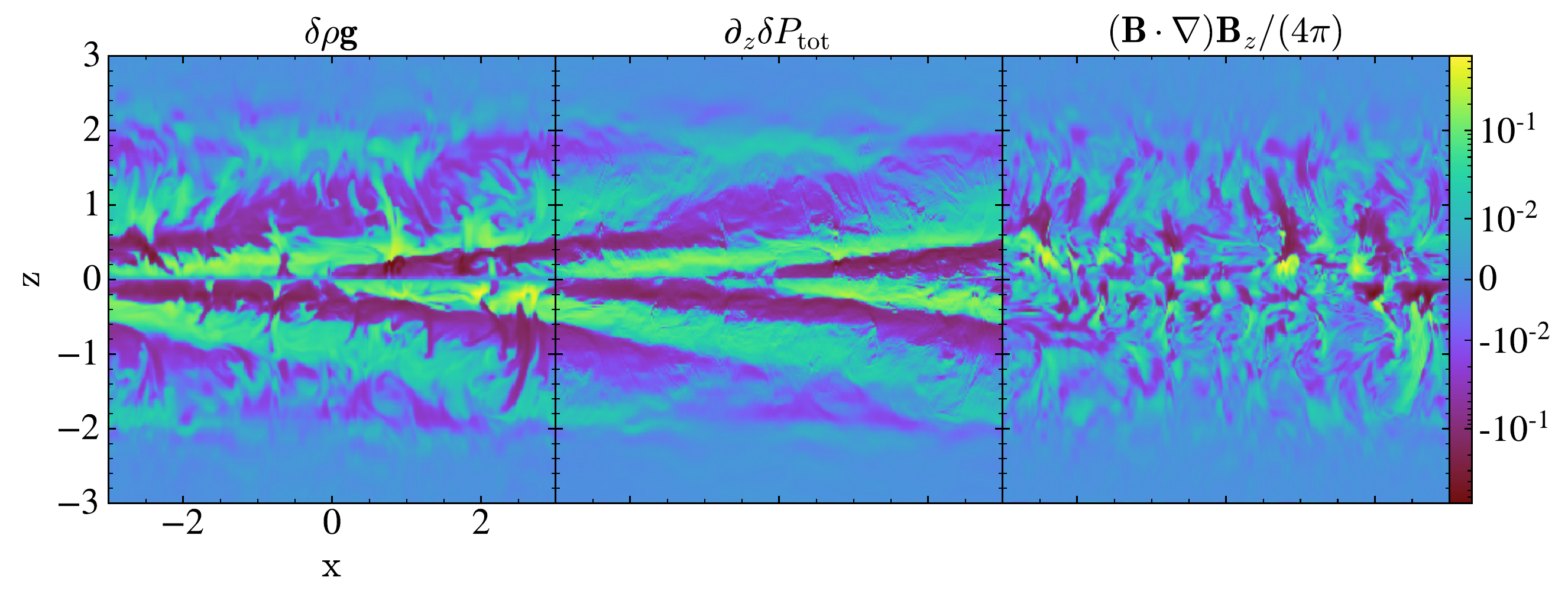}
      \caption{$\beta_0=278$, vertical field}
      \label{fig:tension_by}
    \end{subfigure}
    \begin{subfigure}[b]{\textwidth}
      \centering
      \includegraphics[width=\textwidth]{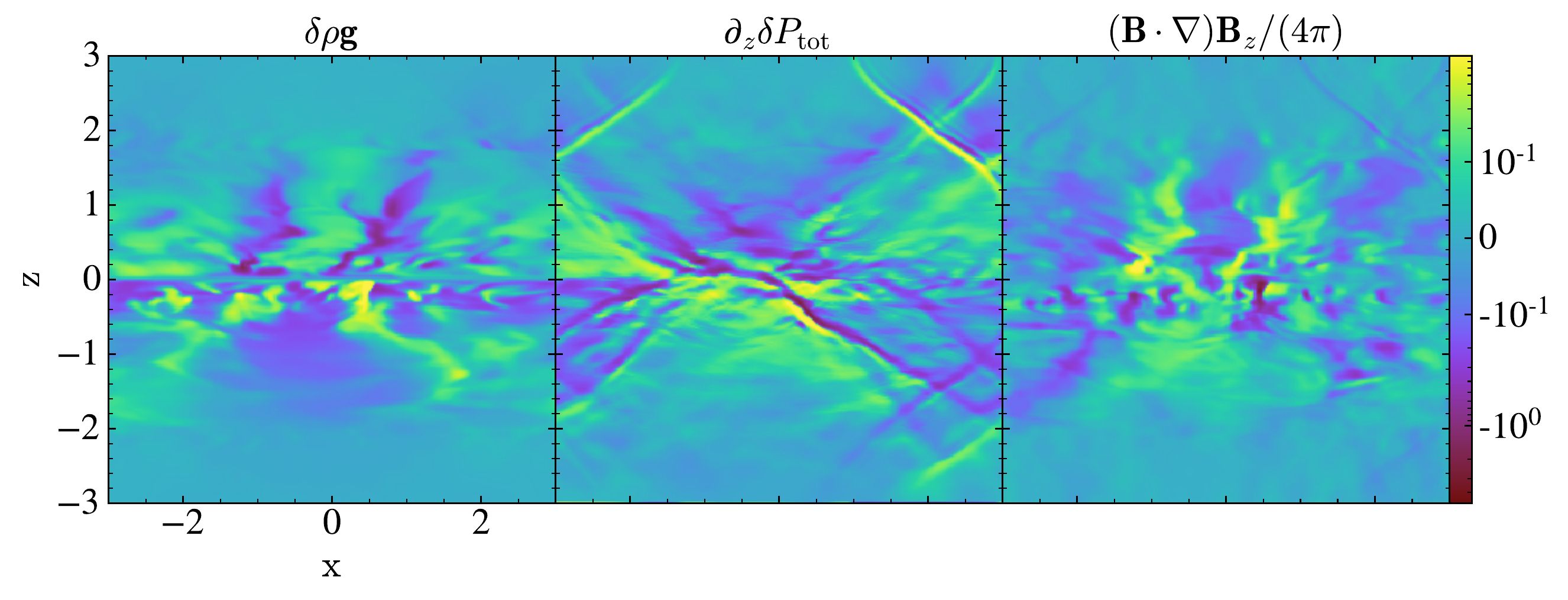}
      \caption{$\beta_0=3$, horizontal field}
      \label{fig:tension_bx_lowbeta}
    \end{subfigure}
  \end{figure*}

  \begin{figure*}
    \ContinuedFloat
    \begin{subfigure}[b]{\textwidth}
      \centering
      \includegraphics[width=\textwidth]{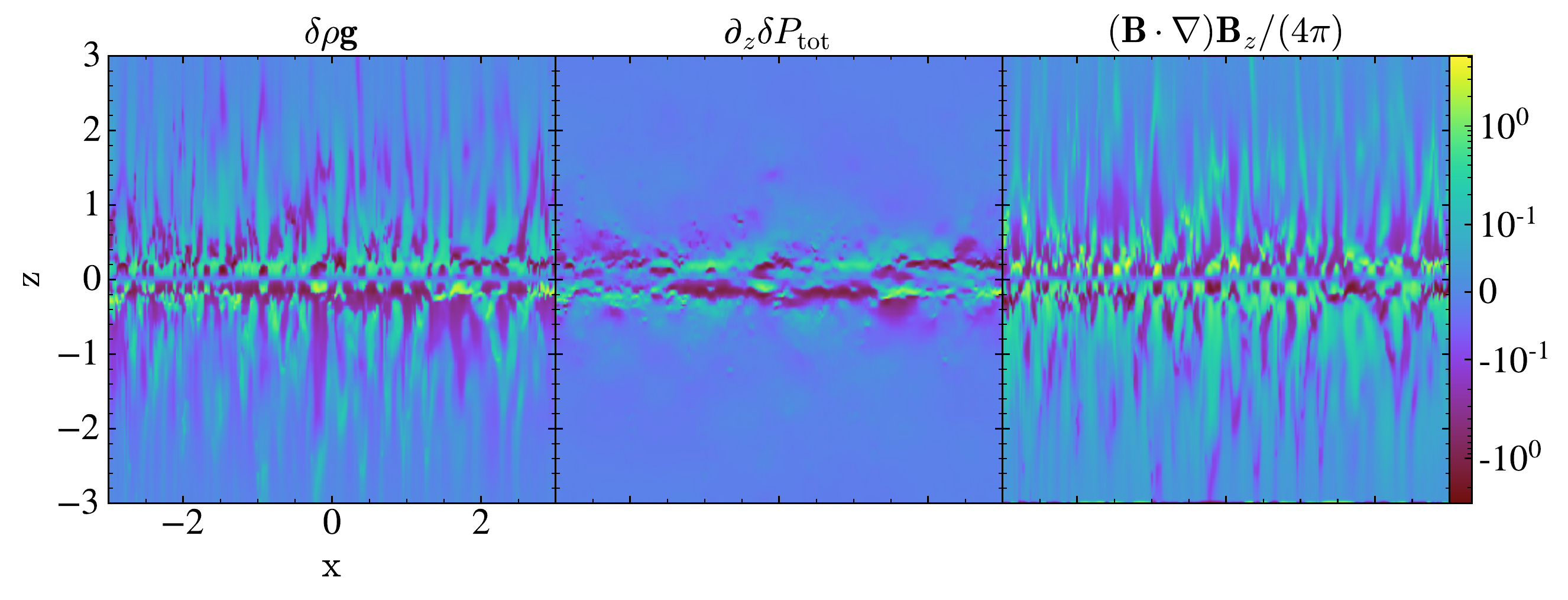}
      \caption{$\beta_0=3$, vertical field}
      \label{fig:tension_by_lowbeta}
    \end{subfigure}
    \caption{Slice plots of the terms of over-density gravity, total pressure gradient and magnetic tension.}
    \label{fig:tension}
  \end{figure*}

  Thus, the existence of a characteristic length scale $l_\mathrm{A}^\mathrm{cool} \sim v_\mathrm{A} t_\mathrm{cool}$ can explain $3$ interesting features seen in our simulations: the dependence of filament size with $\beta$ and $t_\mathrm{cool}/t_\mathrm{ff}$; the existence of a minimum box size for converged results; and mostly importantly, the scaling of RMS density fluctuations with $\beta$.

\subsection{Impact of magnetic field orientation}
\label{sect:orientation}

  In the preceding subsection, we made the case that magnetic field tension damps buoyant oscillations by allowing for magnetic support of overdense gas. Naively, one might expect that this effect will vanish for vertical fields, and thermal instability will no longer be enhanced. Instead, we saw that the amplitude of density fluctuations was roughly independent of field orientation, at fixed $\beta$ and $t_\mathrm{cool}/t_\mathrm{ff}$. Why is this so?

  In Fig. \ref{fig:part_beta278_by_cluster_by} and Fig. \ref{fig:part_beta3_by_cluster_by}, we show particle trajectories for the $\beta_0=278, 3$ cases with vertical fields and $t_\mathrm{cool}/t_\mathrm{ff} = 5.7$. In both cases, buoyant oscillations are suppressed, with amplitudes comparable to the horizontal field case (compare with Fig. \ref{fig:part_beta278_cluster} and Fig. \ref{fig:part_beta3_cluster}). However, the spatial anisotropy of motions appears to be reversed: motions are horizontally (vertically) biased for the weak (strong) field case respectively, opposite to what is seen in the horizontal field case. This squares with the horizontal (vertical) orientation of overdense filaments in the weak (strong) vertical field case (Fig. \ref{fig:slice_deltarho_by}), compared to the opposite orientation in the horizontal field case (Fig. \ref{fig:deltarho_beta278_cluster} and Fig. \ref{fig:deltarho_beta3_cluster}).

  Why are buoyant oscillations suppressed, even for vertical fields? Consider an overdense blob falling under the influence of gravity. Ram pressure leads to a high pressure region upstream of the blob. In the hydrodynamic (and horizontal field) case, this high pressure region is dispersed by lateral gas flows, where unlike in the vertical direction, there are no restoring forces. However, in the vertical field case, horizontal motions are restrained by magnetic tension forces. The high pressure region expands, bending the field lines until the horizontal pressure gradients balance magnetic tension forces. We have verified this force balance explicitly in our simulations. However, this now implies (i) vertical pressure gradients (ii) bent field lines offset from vertical which can exert magnetic tension forces. Both of these can support overdense blobs against gravity and thereby suppress buoyant oscillations.

  We can write the vertical and horizontal force balance equations as:
  \begin{gather}
    \partial_z \delta P_\mathrm{tot} + \frac{(\bm{B} \cdot \nabla) B_z}{4\pi} = \delta \rho g \label{eq:balance_z} \\
    \partial_\perp \delta P_\mathrm{tot} = \frac{(\bm{B} \cdot \nabla) B_\perp}{4\pi}, \label{eq:balance_perp}
  \end{gather}
  where $\delta P_\mathrm{tot}$ is the total hydromagnetic pressure perturbation. To order of magnitude, $B_\perp^2/ 8\pi \sim \delta P_\mathrm{tot}$. Thus, $\delta_z \delta P_\mathrm{tot} \sim \delta_z (B_\perp^2/8\pi)$ and Eq. \eqref{eq:balance_z} can be written as
  \begin{align}
    \frac{(\bm{B}\cdot \nabla) (B_\perp+B_z)}{4\pi} = \frac{(\bm{B}\cdot \nabla) B}{4\pi} = \delta\rho g,
  \end{align}
  which is identical to the force balance equation for horizontal fields. This is why $\delta\rho/\rho$ is similar for both horizontal and vertical fields: for the vertical field case, the lateral confinement of the over-pressured region associated with a falling over-density creates vertical stresses which are comparable to those if field were horizontal, and which suppress buoyant oscillations. This also implies that a similar length scale $l_\mathrm{A}^\mathrm{cool} \sim v_\mathrm{A} t_\mathrm{cool}$ is also singled out.

  We can also understand the change in morphology of the filaments as $\beta$ decreases. When $\beta$ is high, lateral confinement is weak, and gas motions are largely horizontal due to the strong stratification imposed by buoyancy forces, similar to the hydrodynamic case. Thus, filaments are mostly horizontal (Fig. \ref{fig:deltarho_beta278_cluster_by}). However, for strong magnetic fields (low $\beta$), an over-pressured region remains laterally confined, and instead expands in the vertical direction. Thus, filaments are largely vertical (Fig. \ref{fig:deltarho_beta3_cluster_by}). One can think of the magnetic fields as semi-rigid walls, across which gas is not free to flow. Since horizontal pressure balance is no longer required, alternating strips of high and low pressure gas develop, corresponding to high and low density strips respectively, each in a separate state of hydrostatic equilibrium.

  In Fig. \ref{fig:tension_by}, we show the perturbed forces for the vertical field case. We see that for high $\beta$, $\partial_z \delta P_\mathrm{tot}$ largely matches $\delta\rho g$. Quantitatively, perturbed pressure gradient dominate vertical magnetic tension support by a factor of $5$. The field lines are bent to roughly horizontal around overdense regions, and the influence of magnetic tension is indirect (by confining over-pressured regions). By contrast, for $\beta=3$ (Fig. \ref{fig:tension_by_lowbeta}), we see that the magnetic tension clearly provides the support. Field lines are only slightly perturbed from vertical, and magnetic tension support is direct.

\subsection{Independence to cooling curve}
\label{sect:independence_cooling_curve}

  As previously discussed in \S\ref{sect:cooling_curve}, the amplitude of density fluctuations is independent of the cooling curve, and is completely determined by $\beta$ and $t_\mathrm{cool}/t_\mathrm{ff}$, as in Eq. \eqref{eq:delrho_time}. These two parameters specify $l_\mathrm{A}^\mathrm{cool} \sim v_\mathrm{A} t_\mathrm{cool} \sim H/\sqrt{\beta} (t_\mathrm{cool}/t_\mathrm{ff})$ in the hot medium, which sets the balance between cooling and magnetic stresses in the hot medium. The interesting exception for volumetric heating (where $\Lambda \propto T^\alpha$) is the apparent independence of $\delta\rho/\rho$ to $t_\mathrm{cool}/t_\mathrm{ff}$ at low $\beta$. As mentioned in \S\ref{sect:cooling_curve}, this is a box size artifact; RMS density fluctuation drops close to the $(t_\mathrm{cool}/t_\mathrm{ff})^{-1}$ scaling in a larger box.

  Interestingly, this only happens for the galaxy cooling curve with horizontal fields at low $\beta$. Examination of the box shows a single dominant filament with a highly rarefied and overheated hot medium, with a significantly long cooling time such that $l_\mathrm{A}^\mathrm{cool} \sim v_\mathrm{A} t_\mathrm{cool} > L_\mathrm{box}$.\footnote{Note that in the initial hot medium, $l_\mathrm{A}^\mathrm{cool} \sim 1.4 H < L_\mathrm{box}$; $l_\mathrm{A}^\mathrm{cool}$ subsequently increases, largely because of the increase in cooling time.} The density and entropy PDFs become bimodal, indicating genuine fragmentation into a multi-phase medium. By contrast, in the larger box with identical $t_\mathrm{cool}/t_\mathrm{ff}$ and $\beta$, the hot medium does not evolve significantly, with $l_\mathrm{A}^\mathrm{cool} < L_\mathrm{box}$; therefore, the entropy and density PDFs remain unimodal.

  We conclude that in a system with sufficient separation of scales, then the saturated state of thermal instability obeys the scaling of $\delta\rho/\rho \propto \beta^{-1/2} (t_\mathrm{cool}/t_\mathrm{ff})^{-1}$ (Eq. \eqref{eq:delrho_time}). However, systems where $l_\mathrm{A}^\mathrm{cool} \gsim H$ are prone to ``overcooling''.

  Under isobaric conditions, emissivity fluctuations scales as $\delta \epsilon/\epsilon \propto (\alpha-2) \delta T/T$, implying stronger emissivity fluctuations for the galaxy ($\alpha=-1$) as opposed to cluster ($\alpha=0.5$) cooling curves. In a finite box, the former is prone to a thermal runaway where $l_\mathrm{A}^\mathrm{cool}$ grows beyond the box size and all available modes are destabilized by magnetic fields.

  We speculate that this apparent box size effect may be of direct physical relevance to galaxy halos. It is common to find $\beta \sim \ \text{few}$, $t_\mathrm{cool}/t_\mathrm{ff} \gsim 10$ in such halos. By the usual hydrodynamic criteria $t_\mathrm{cool}/t_\mathrm{ff} < 10$ \citep{sharma12}, such halo gas should be thermally stable. With the addition of magnetic fields, not only is such a configuration thermally unstable, but because $l_\mathrm{A}^\mathrm{cool} \gsim H$ (and even $l_\mathrm{A}^\mathrm{cool} \gsim r$), it can becomes violently thermal unstable, approaching a state where all large scale modes are thermally unstable and gravity drops out of the problem. Essentially, with strong magnetic fields, thermal instability is no longer quenched by non-linearity induced by falling overdense blobs, and proceeds with the same vigor as in an unstratified medium.

  This has obvious implications for manufacturing cold gas in halo outskirts, where it is seen (e.g. in Ly$\alpha$ florescence; \citealt{borisova16}), and in circumstances where thermal instability would otherwise not be an obvious culprit.

\section{Discussion}
\label{sect:discussion}

\subsection{Missing physics}

  In this first study, we have necessarily made some simplification or left out some ingredients. These include:

  \begin{enumerate}

    \item {\it Thermal conduction} Thermal conduction quenches thermal instability parallel to the magnetic field on scales below the Field length, $\lambda_\mathrm{F} \approx \sqrt{\kappa(T) T / n^2 \Lambda(T)}$, where $\kappa(T) \approx n_\mathrm{e} k_\mathrm{B} v_\mathrm{e} \lambda_\mathrm{e}$ is the conduction coefficient, and $v_\mathrm{e}$, $\lambda_\mathrm{e}$ are the electron thermal velocity and mean free path respectively. If we compare this to the characteristic scale $\lambda_\mathrm{A}^\mathrm{cool}$ we have found for thermal instability, we get
    \begin{align}
      \left(\frac{\lambda_\mathrm{F}}{\lambda_\mathrm{A}^\mathrm{cool}}\right)^2
        &=\left(\frac{c_\mathrm{s}}{v_\mathrm{A}}\right)^2 \frac{m_\mathrm{p}}{m_\mathrm{e}} \frac{\lambda_\mathrm{e}}{c_\mathrm{s} t_\mathrm{cool}} \notag \\
        &=0.02 f \beta_3 \Lambda_{-22} T_6^{1/2},
    \end{align}
    where $f$ is a suppression factor accounting for reduction in conduction below the \emph{Spitzer} rate, due perhaps to small scale magnetic field tangling or electron scattering by plasma instabilities, and $\beta_3 \equiv \beta/3$, $\Lambda_{-22} = (\Lambda(T) / 10^{-22})\ \mathrm{erg\ s^{-1}\ cm^3}$, and $T_6 = (T/10^6)\ \mathrm{K}$. Since $\lambda_\mathrm{F} < \lambda_\mathrm{A}^\mathrm{cool}$, thermal conduction does not obviously play a significant role, although the separation of scales is not very large and the true impact of conduction will have to be assessed in simulations with field-aligned conduction.\footnote{Previous such simulations have found little difference in simulations with and without anisotropic conduction \citep{mccourt12}. Note that this simulations were initialized with \emph{extremely} weak fields ($\beta\sim 10^6$), so they did not see the dynamical effects we report here. In addition, \citet{wagh14} employed anisotropic thermal conduction and tested different field geometry including tangled fields, finding little difference between hydro and MHD cases as well. This is most likely because the field strength is also weak ($\beta\sim10^3$), and $v_\mathrm{A} t_\mathrm{cool}$ is much smaller than the length scales discussed here.} Note that the suppression factor $f$ is unknown and potentially large; there is no evidence for thermal conduction anywhere in dilute plasmas apart from the solar wind, which has a large scale, ordered magnetic field \citep{bale2013electron}.

    \item {\it Geometry} We have only simulated planar geometry, and the impact of spherical geometry could be important, once $\lambda_\mathrm{A}^\mathrm{cool}$ becomes large and approaches the radius $r$. Indeed, once $\lambda_\mathrm{A}^\mathrm{cool}/r \sim \lambda_\mathrm{A}^\mathrm{cool}/H \sim (t_\mathrm{cool}/t_\mathrm{ff}) \beta^{-1/2} > 1$, we might expect magnetic fields to destabilize all available modes, similar to the ``over-cooling effect'' seen in small boxes with planar geometry (see \S\ref{sec:box_size}). This will have to be explored in future work. Note that in the hydrodynamic case, spherical geometry was previously thought to boost the threshold value of $t_\mathrm{cool}/t_\mathrm{ff}$ required for a multi-phase medium to develop due to differing compression on a sinking fluid element \citep{sharma12}, but it has subsequently been shown that once an apples-to-apples comparison of where the cold gas appears is performed, there is little to no difference between spherical and Cartesian geometry \citep{choudhury2016cold}. The latter authors found instead that the amount of cold gas depends on how $t_\mathrm{cool}/t_\mathrm{ff}$ varies with radius.

    \item {\it Magnetic field topology} We have seen that straight horizontal and vertical magnetic fields have comparable effects in enhancing thermal instability. This suggests that magnetic field orientation does not play a significant role and that a tangled field would give similar results, although the case $\lambda_\mathrm{B} < \lambda_\mathrm{A}^\mathrm{cool}$ (where $\lambda_\mathrm{B}$ is the magnetic coherence length) deserves more careful study. It may well be that the characteristic length scale should be $\mathrm{min} (\lambda_\mathrm{A}, \lambda_\mathrm{A}^\mathrm{cool}, \lambda_\mathrm{B})$.

    \item {\it Hydrostatic and thermal equilibrium} Our model is by construction always in global hydrostatic and thermal equilibrium. While this is clearly an excellent approximation for galaxy groups and clusters, it is potentially more questionable at galaxy scales, particularly for lower mass ``cold mode'' halos which may lack a hot hydrostatic halo, or in systems where galactic winds drive a strong outflow. While local thermal instability is suppressed in a cooling flow, even weak magnetic fields enable thermal instability to once again operate \citep{loewenstein90,balbus91,hattori1995non}. It would be interesting to study if magnetic fields can play a similar role in modifying thermal instability in galactic winds. Thermal instability in galactic winds has been the subject of much recent interest as a means of explaining cold gas clouds outflowing at high velocity; since they are born at high velocity, they can evade the destruction by Kelvin-Helmholtz instabilities which would take place if they were accelerated by the wind \citep{martin15,thompson16,scannapieco17}.

    \item {\it Large density perturbations} Our simulations start with tiny density perturbations, from which thermal instability develops and is independent of a sufficiently small magnitude of initial perturbations. However, the density perturbations in galaxy halos might be potentially large due to strong turbulence or galactic winds. \citet{pizzolato05} and \citet{singh15} suggest that thermal instability can develop even at large $t_\mathrm{cool}/t_\mathrm{ff}$ when initial density perturbations are large, but this regime has not been explored in our study. Therefore, it is necessary in future studies to relax the assumptions about hydrostatic equilibrium by including driven turbulence into simulations, and investigate how initially large density perturbations produced from strong turbulence can modify the evolution of thermal instability in MHD cases.

  \end{enumerate}

\subsection{Conclusions}

  In this paper, we explore the effects of magnetic fields on thermal instability in a stratified medium with 3D MHD simulations. Astrophysical plasmas are invariably magnetized and our main message is that this is a regime where pure hydro is clearly insufficient; MHD effects simply cannot be swept under the rug. The main application we have in mind is the formation of cold gas in galaxy ($\beta\sim\ \text{few}$) and cluster ($\beta\sim50$) halos, but these results should be broadly applicable to other environments. We find that magnetic fields have a striking impact on both the amplitude and morphology of thermal instability, even at surprisingly low magnetic field strengths. This greatly broadens the radial range in which halo gas can be thermally unstable. It is a potentially promising explanation for the ubiquity of cold gas in absorption line spectra of the CGM.

  Our main results are as followings:

  \begin{enumerate}

    \item Magnetic fields have a strong influence in enhancing the \emph{amplitude} of density fluctuations due to thermal instability, and its ability to fragment into multi-phase medium This enhancement appears even at very weak magnetic fields, and the MHD results match the hydrodynamic results only at $\beta\sim 1000$. Otherwise, $\delta\rho/\rho \propto \beta^{-1/2}$, where $\beta\equiv P_\mathrm{gas}/P_\mathrm{B}$ [see equation \eqref{eq:delrho_time}]. This enhancement arises because MHD stresses weaken the effective gravitational field, and thus reduce the magnitude of buoyant oscillations (which otherwise disrupt condensing cold gas) by supporting overdense gas against gravity. This can be seen in both particle tracer plots and slice plots of linearized perturbed forces.

    \item Magnetic fields have a strong influence on the \emph{morphology} of cold gas filaments. The strength and orientation of magnetic fields strongly influence the size and orientation of cold filaments, even for very weak fields, which can be understood in terms of the MHD stresses. In particular, the spacing between filaments scales with the characteristic scale $l_\mathrm{A}^\mathrm{cool} = v_\mathrm{A} t_\mathrm{cool}$ (see below).

    \item Thermal instability in the presence of magnetic fields imprints a characteristic scale $l_\mathrm{A}^\mathrm{cool} = v_\mathrm{A} t_\mathrm{cool}$ on cooling gas. This scale is evaluated for conditions in the \emph{hot medium} (unlike the shattering length $c_\mathrm{s} t_\mathrm{cool}$, which is evaluated for conditions in the cool medium; \citealt{mccourt18}), since it determines the characteristic scale in the hot medium at which buoyant oscillations are suppressed and gas cooling is enhanced. The appearance of this characteristic scale can explain three features in our simulations: the scaling of density fluctuations $\delta\rho/\rho \propto \beta^{-1/2}$ [by setting the the scale on which magnetic tension operates; $B^2/\lambda_\mathrm{A}^\mathrm{cool} \sim \delta\rho g$, Eq. \eqref{eq:grav_tension}], the scaling of filament sizes with $\beta$ and $t_\mathrm{cool}/t_\mathrm{ff}$ in simulations, and the ``over-cooling'' effect seen when the box size is smaller than this scale, implying that all modes in the box are destabilized by magnetic fields.

    \item To zeroth order, the amplitude of density fluctuations and cold gas fraction are independent of magnetic field orientation and cooling curve shape. The independence from field orientation arises because magnetic tension can be shown to either support overdense gas directly (for initially horizontal fields) or indirectly (by confining pressurized hot gas, which in term supports the overdense gas or tilts the field lines so that magnetic tension support is important). The independence from cooling curve arises because the effect of magnetic fields essentially depends on $\sim B^2/\lambda_\mathrm{A}^\mathrm{cool}$, where $\lambda_\mathrm{A}^\mathrm{cool}$ is evaluated in the hot phase, not the cooler phase. An important exception is the case when mass dropout is sufficiently large that the properties of the hot phase evolve; thus, density fluctuations in low $\beta$, galaxy cooling curve (which is more unstable to fragmentation to multi-phase gas) requires a larger box (to avoid ``over-cooling'') for converged results.

  \end{enumerate}

  We close by noting that while magnetic fields clearly enhance thermal instability in a stratified medium, there are also other mechanisms for doing so. Any process that damps buoyant oscillation can in principle enhance thermal instability. Thus, for instance, a flattening of the entropy gradient (Voit et al. 2017), which weakens buoyant restoring forces, and also expel low entropy gas to large radii (where heating is weak), can potentially have similar effects, though quantifying the impact of these effects requires detailed numerical simulation. In the MHD case, MHD stress oppose the bending of field lines and act as a drag force on an oscillating  blob (similar to how the drag force on a moving blob is enhanced by magnetic fields; \citealt{mccourt15}), thus damping buoyant oscillations. The key feature we find that make it particularly conclusive for enhancing thermal instability is that this mechanism becomes important at remarkably low field strengths, and independent of field orientation. This implies that it could potentially be important throughout the halo. Furthermore, magnetic fields are ubiquitous and long-lived. By contrast, other processes are less universal and either spatially or temporally confined. Halos are generally strong stratified by entropy and exhibit flat entropy profiles only in their cores, if at all. Turbulence decays in an eddy turnover time, and the strong supersonic turbulence needed to overcome buoyancy restoring forces is only present in halos during periods of strong AGN or starburst activity. The effects of magnetic fields are much more universal, and we ignore them at our peril.

\section{Acknowledgements}

  We thank Mark Voit for helpful conversations. We also thank our referee, Prateek Sharma, for helpful and insightful comments which improved the paper. We acknowledge NASA grants NNX15AK81G, NNX17AK58G and HST-AR-14307.001-A for support. We also acknowledge support from the University of California Office of the President Multicampus Research Programs and Initiatives through award MR-15-328388. This research has used the Extreme Science and Engineering Discovery Environment (XSEDE allocations TG-AST140086). We have made use of NASA's Astrophysics Data System and the yt astrophysics analysis software suite \citep{turk2010yt}. SPO thanks the law offices of May Oh and Wee for hospitality during the completion of this paper. 

\bibliographystyle{mnras}
\bibliography{master_references}

\end{CJK}
\end{document}